# Benchmarking of Oxygen Adsorption using TPD Spectroscopy for Accurate DFT Prediction of ORR on Anatase Titanium Dioxide (101)

Shibghatullah Muhammady,[1,†,*] Jun Haruyama,[1,†] Shusuke Kasamatsu,[2] Osamu Sugino[1,*]

[1]*The Institute for Solid State Physics, The University of Tokyo, 5-1-5 Kashiwanoha, Kashiwa, Chiba 277-8581, Japan*

[2]*Academic Assembly (Faculty of Science), Yamagata University, 1-4-12 Kojirakawa, Yamagata, Yamagata 990-8560, Japan*

[†]*Present address: Solid State Chemistry Laboratory, Pioneering Research Institute, RIKEN, 2-1 Hirosawa, Wako, Saitama, 351-0198, Japan*

*E-mails: shibghatullah.muhammady@riken.jp, sugino@issp.u-tokyo.ac.jp

**Abstract**

Despite the wide use of first-principles calculations to elucidate catalytic reaction mechanisms, the reliability of the theory remains unknown for reactions that initiate with $O_2$ adsorption, as few papers in the literature have systematically verified the accuracy of the calculations. Here, we have overcome this issue by comparing calculated $O_2$ adsorption energy ($E_{ads}$) to simulated $E_{ads}$ distribution from experimental temperature-programmed desorption (TPD) spectrum. The distribution obtained based on equilibrium thermodynamic arguments is in good agreement with the calculated $E_{ads}$ from modelling $O_2$ adsorption on *anatase*-$TiO_2$(101) using the generalized gradient approximation and van der Waals density functionals (vdW-DFs) with Hubbard correction, which is particularly the case when vdW-DF3-opt2 is used. On this basis, we concluded that the oxygen reduction reaction (ORR) initiates from a physisorbed $O_2$ thermodynamically as stable as that in the gas phase. In contrast, the calculated $E_{ads}$ for Pt(111) exhibit a significant overestimation of the $O_2$ adsorption due to excessive vdW correction. Therefore, a systematic investigation based on TPD spectroscopy can be used to diagnose the reliability of theoretical predictions of ORR and is expected to lead to improvements in exchange-correlation functionals for catalysts.

## 1. INTRODUCTION

The oxygen reduction reaction (ORR) is a typical heterogeneous catalytic reaction whose activity is strongly correlated with oxygen ($O_2$) adsorption and activation.[1] On typical metal surfaces, the molecularly adsorbed oxygen ($O_2^*$) is categorized into two main states: the least stable physisorbed state, with the adsorption energy ($E_{ads}$) of approximately 0.1 eV, and the chemisorbed state, which exists either in the superoxo ($O_2^-$) or peroxo ($O_2^{2-}$) form ($E_{ads}$ = 0.5 to 2 eV). These molecularly adsorbed states are weakly bound on the surface compared with the atomically adsorbed state ($E_{ads}$ = 1 to 10 eV).[2] We employ the sign convention where positive $E_{ads}$ corresponds to thermodynamically stable adsorption. Furthermore, O_2* is thermodynamically stable on noble metals such as Pt, Pd, Ag, and Au whereas it is dissociatively adsorbed as O* on transition metals such as Ni, Co, Fe, Rh, and Ru. This difference in adsorption tendency is influenced by the charge transfer, depending on the geometric and electronic structures of the surfaces.[2] In the former case (noble metals), the ORR is expected to proceed with the associative mechanism, beginning with $O_2^* + (H^+ + e^-) \rightarrow O_2H^*$, while in the latter case (transition metals), it proceeds with the dissociative mechanism, starting from $O^* + (H^+ + e^-) \rightarrow OH^*$. In either scenario, the mechanism is delicately influenced by the chemical environment described by, for example, coverage and electrode potential ($\Phi$).[3,4] Moreover, oxide catalysts, which have garnered significant attention in recent years as stable and inexpensive alternatives to Pt catalysts for fuel cells, introduce further complexity to the adsorption chemistry through various local structures involving anions, cations, as well as vacancy and substitutional defects.[5–7] Therefore, a crucial first step in ORR catalyst design is to thoroughly understand the adsorption state, for which the first-principles calculation based on the density functional theory (DFT) can play an important role.

Quantitatively predicting oxygen adsorption properties is quite challenging, as is evidenced by the wide scatter in reported $E_{ads}$ values for the prototypical Pt(111) surface. Previous studies using the GGA-PBE functional[8,9] have reported $E_{ads}$ values ranging from 0.34 to 0.79 eV for molecular chemisorption and 0.99 to 2.4 eV for atomic adsorption (see Table S4 in the Supporting Information).[10–16] The scatter is not simply due to different adsorption geometries considered by various authors. Possible reasons include inaccurate pseudopotentials, variations in the number of Pt layers in the surface slab model, differences in how surface atoms are relaxed, the specific bulk lattice constant used, and/or insufficient k-point and plane wave cutoff convergence. Some authors have reported careful convergence studies for hydrogen on Pt(111), for instance.[17] In any case, such significant scatter, even when employing the same density functional approximation, highlights the technical difficulties in obtaining reliable energy values with errors of less than approximately 0.1 eV, which is crucial for accurate surface science.

Another issue is the lack of a computationally feasible density functional approximation that is sufficiently accurate for all adsorption chemistries. The difficulty starts with the description of the spin-triplet ground state $^3O_2$[18] of the isolated molecule. Conventional semilocal exchange-correlation (XC) functionals significantly overestimate the O–O binding energy: by as much as approximately 2 eV when employing the local density approximation (LDA), approximately 1 eV with generalized gradient approximations (GGAs), and approximately 0.5 eV with meta-GGAs.[19] This overestimation persists even with more modern hybrid functionals and van der Waals density functionals (vdW-DFs).[20] This overbinding error most seriously affects the $E_{ads}$ of dissociatively adsorbed O*, and it is often corrected using the experimental binding energy of isolated $O_2$.[21]

Describing the molecularly adsorbed $O_2$* presents a different set of challenges. The molecule-surface bonding character can vary significantly depending on the catalyst surface, ranging from a predominantly van der Waals (vdW) type to a mostly covalent one. It is important to note that even when comparing among well-tested GGAs, the results are even qualitatively different. For example, studies comparing the GGA-RPBE functional to the GGA-PW91 functional report a rather weak molecular $E_{ads}$ of 0.1 eV for the former and a much stronger 0.6 eV for the latter.[22,23] This highlights the uncertainties of conventional density functional approximations for molecular adsorption. A significant computational advancement for this situation is random phase approximation (RPA), which minimizes GGA self-interaction errors while also covering long-range non-local correlations based on many-body perturbation theory.[24–26] However, RPA is currently too computationally expensive, making it unfeasible to apply to complex systems.[27] This computational cost is also a limitation for hybrid functionals.[2] One promising approach to address these challenges is the nonlocal van der Waals density functional (vdW-DF). By simplifying the XC term within the RPA,[28] the vdW-DFs offer improved performance for $O_2$ adsorption compared to semilocal functionals, while maintaining a comparable computational cost.[29–32] However, application to ORR intermediates on various surfaces is still rather limited.

The fact that $O_2$* often carries electron spin introduces an additional layer of complexity. Weak van der Waals (vdW) interactions and spin-spin interactions are similar in magnitude and may compete, meaning that an accurate description of both might be necessary for precise predictions of the bonding structure and, consequently, the $E_{ads}$. In this context, researchers have observed that vdW-DFs significantly overestimate the interactions between $O_2$ molecules with opposite spin.[33] However, studies have shown that applying the Hubbard-$U$ correction on the O 2p orbital alleviates this issue, enabling the correct reproduction of the ground state monoclinic structure of solid oxygen.[34,35] It is important to note that the magnetic moment of $O_2$* is surface-dependent, so a simple, universal correction cannot be applied across all surfaces.

For instance, the superoxo and peroxo on Pt(111) exhibit finite and zero magnetic moments, respectively.[2]

Given the theoretical uncertainties highlighted above, careful verification of each DFT calculation is crucial at this stage. We contend that an appropriate way to address this need is through comparison with experimental temperature-programmed desorption (TPD) spectra. TPD spectroscopy is a powerful experimental technique for examining adsorption characteristics. It measures the number of desorbed molecules from a surface as the temperature is increased, and the results are usually interpreted using the Polanyi-Wigner equation:

$$-\frac{d\theta}{dt} = \theta^n \nu_0 e^{-(E_{ads}/k_B T)} \quad (1)$$

where $\theta$, $n$, $\nu_0$, $k_B$, and $T$ are the coverage, desorption order, pre-exponential factor, Boltzmann constant, and absolute temperature, respectively. $E_{ads}$ represents the desorption barrier or the energy difference between the gas and adsorbed phases. For example, TPD spectra of molecular ($O_2^* \rightarrow O_2(g)$) and atomic ($O^* \rightarrow \frac{1}{2}O_2(g)$) desorption of $O_2$ from Pt(111) have been used to estimate their respective $E_{ads}$ values as 0.35 – 0.38 and 1.78 – 1.97 eV.[36,37] It is also possible to measure oxygen adsorption on oxide catalysts, which are potential replacements for expensive platinum catalysts.[38,39] Notably, for $O_2$ adsorption on *anatase*-TiO$_2$(101), the estimated $E_{ads}$ is 0.19 ± 0.02 eV.[38] The spectroscopy has also proven effective for understanding surface functional groups,[40] determining the optimum temperature for catalytic reactions,[41] and identifying the nature and strength of active sites on catalyst surfaces.[42]

Extracting $E_{ads}$ from experimental TPD spectra using Equation (1) requires knowing the value of the pre-exponential factor ($\nu_0$), which is often difficult to obtain from either experiment or theory. If $\nu_0$ is assumed to be constant, a further difficulty arises when applying this equation to surfaces with varying $E_{ads}$ values, leading to overlapping TPD peaks. Consequently, the reported $E_{ads}$ values from TPD experiments often carry significant uncertainties.[43–45] Conversely, recent research has proposed an alternative method based on the equilibrium thermodynamics of an ideal lattice gas. This approach is applicable to a surface with multiple types of $E_{ads}$, providing the distribution of adsorption energies ($\rho(E_{ads})$), and inverse analysis to extract $\rho(E_{ads})$ from experimental TPD spectra at saturation coverage is also possible.[46] This achievement has enabled direct comparison of TPD spectra or $\rho(E_{ads})$ between experiment and theory, thus facilitating atomistic interpretation of the experimental results and evaluation of exchange-correlation functional accuracy.

In this study, we carefully investigate the accuracy of density functional approximations for describing the adsorption of molecular ($O_2$*) and atomic (O*) oxygen on Pt(111), a typical ORR catalyst, and anatase-type titanium oxide (*anatase*-$TiO_2$(101)), which has recently gained attention as a promising alternative ORR catalyst. We reanalyze available experimental TPD data using the recently developed methodology based on equilibrium thermodynamics[46] to obtain $\rho(E_{ads})$. This $\rho(E_{ads})$ is compared with $E_{ads}$ values derived from the Polanyi-Wigner equation (see Equation 1). The extracted $\rho(E_{ads})$ data are also compared to predictions using GGA-PBE and vdW-DF functionals, both with and without Hubbard $U$ correction, considering various adsorption geometries. While a similar comparative study on DFT functionals exists for Pt(111), it does not cover the molecular adsorption.[31] For *anatase*-$TiO_2$(101), we found that $\rho(E_{ads})$ values were well reproduced by calculations using vdW-DFs. This highlights the importance of vdW corrections, which have been largely overlooked so far, and, more importantly, the possibility of obtaining reliable free energy curves for ORR. Building on this, we also evaluated the reliability of ORR free energy predictions. In contrast, for Pt(111), insufficient agreement was found with experimental results, indicating that reliable theoretical predictions remain challenging. Based on these findings, while popular XC functionals can only partially cover the range of materials, it is now possible to predict the ORR mechanism by first confirming $E_{ads}$ of $O_2$. This represents an important advancement in catalyst research.

## 2. RESEARCH METHODOLOGY

**2.1. Extraction of $\rho(E_{ads})$ from experimental TPD data.** From experimental TPD spectra,[36,38] $\rho(E_{ads})$ was determined by a deconvolution-like technique based on equilibrium thermodynamics and the Langmuir adsorption model.[46] The extraction procedure requires as input a TPD spectrum starting from saturation coverage, the temperature ramp rate, the gas species (in this case, $O_2$), the number of adsorption sites per unit area (i.e., the number of adsorbates desorbing in the TPD measurement), and the sticking probability of impinging molecules.

For the Pt (111) surface, we analyze the experimental TPD spectrum with a temperature ramp rate of 4.75 K/s after saturated $O_2$ exposure of approximately $10.60 \times 10^{14}$ molecules/cm$^2$ at 100 K [Figures 1(a) and 1(b)].[36] It has been shown that predominantly molecular adsorption occurs at 100 K with an estimated $O_2$* saturation coverage of about $6.1 \times 10^{14}$ molecules/cm$^2$; upon heating, molecular desorption occurs as seen in the sharp peak in the 150-200 K range of the TPD [Figure 1(a)]. Only about $4.2 \times 10^{14}$ molecules/cm$^2$ desorb in this temperature range, and the remaining molecules dissociate at above 170 K and form a (2 × 2) overlayer of atomic adsorbates corresponding to O* coverage of $3.8 \times 10^{14}$ atoms/cm$^2$. Upon further heating, all these adsorbates recombine to desorb as oxygen molecules corresponding to the broad peak in the 600-900 K range [Figure 1(b)].

For O$_2$* adsorption on *anatase*-TiO$_2$(101), the TPD spectrum after an exposure of 3.2 Langmuir was used. The temperature ramp rate was 1 K/s. Two prominent peaks appear at 34 K and 64 K, where the higher temperature peak has been interpreted as originating from saturated monolayer adsorption and the lower temperature peak from additional multilayer adsorption.[38] Since the TPD analysis requires input TPD spectra with saturated peaks only, we extract the monolayer adsorption peak assuming exponential decay in the pre-peak region as shown in Figure 1(c). The saturated monolayer coverage is taken to be $5.16 \times 10^{14}$ cm$^{-2}$ based on the experimental lattice parameters,[47] corresponding to one O$_2$* per 5-coordinated Ti atom (Ti$_{5c}$).[38] For the sticking probability of impinging adsorbates, we assume the following Langmurian form:

$$s = s_0(1 - \theta_{\text{rel}})^N \tag{2}$$

Here, $\theta_{\text{rel}}$ is the coverage divided by the saturation coverage and $s_0$ represents the initial sticking coefficients onto clean surfaces, with $N = 1$ for molecular adsorption and $N = 2$ for atomic adsorption of a diatomic molecule.[48] We adopted $s_0 = 0.05$ for Pt(111)[36] and $s_0 = 0.9$ for *anatase*-TiO$_2$(101), as estimated in the literature. We expect that errors in $s_0$ will not significantly affect the results [42].

**2.2. Calculation methods.** Quantum ESPRESSO code[49–51] was used for spin-polarized DFT calculations[52,53] using four XC functionals, i.e., Perdew-Burke-Ernzerhof functional within the GGA[9] and different vdW-DFs [Section 2.2.], as implemented into the code by Thonhauser group.[54,55] The projector augmented wave[56] method was used for generating the pseudopotentials of all the involved elements. Calculation details, including convergence tests for bulk and surface slab systems, are provided in the Supporting Information (Section S1).

**2.3. Non-local correlation from vdW-DFs.** Briefly overviewing vdW-DFs, the non-local correlation energy functional is formulated by

$$E_c^{nl}[n] = \frac{1}{2}\int d^3\mathbf{r} \int d^3\mathbf{r}' n(\mathbf{r})\phi(\mathbf{r},\mathbf{r}')n(\mathbf{r}') \tag{3}$$

where the kernel $\phi(\mathbf{r}, \mathbf{r}')$ depends on $(\mathbf{r} - \mathbf{r}')$ and the density $n$, connecting different space regions from the adiabatic connection formula. The GGA exchange energy is given by

$$E_x^{\text{GGA}}[n] = \int d^3\mathbf{r} n(\mathbf{r})\varepsilon_x^{\text{hom}}[n(\mathbf{r})]F_x(g) \tag{4}$$

where $\varepsilon_x^{\text{hom}}$ is the exchange per particle covering the homogeneous electron gas, while $F_x(g)$ is the GGA exchange enhancement factor, a function of the reduced gradient $g$.[57,58] Within $F_x(g)$, a factor $Z_{ab}$ in vdW-DF1 and vdW-DF2 was tuned to improve the performance of the XC functional based on the second-order gradient expansion corresponding to the slowly varying electron gas[54,59] and second-order large-$N$ expansion for the neutral atom,[60] respectively. The performance was further improved by tuning the momentum dependence of the plasmon dispersion included in the explicit form $E_c^{nl}[n]$ as the additional degree of freedom within the vdW-DF framework. Therein, the switching factor $h$ is defined as $h(y) = 1 - (1 + \gamma y^2 + \gamma^2 y^4 + \alpha y^8)^{-1}$, where $g$ and $y$ are adjustable parameters.[58] This form alters the previous form formulated as $h(y) = 1 - \exp(-\gamma y^2)$.[57,60] We employed the vdW-DF-ob86[61] and vdW-DF2-B86R,[62,63] which improves the description of geometries and energetics of solids, molecules, and adsorption structures. The third generation of vdW-DF3-opt2[58] is also used.

**2.4. Modeling the surfaces.** Pt(111) and *anatase*-TiO$_2$(101) surface slab structures are used as the computational models for O$_2$ adsorption. We applied a Hubbard U correction of 4 eV to the Ti 3d orbitals in *anatase*-TiO$_2$, following previous studies.[64,65] For comparison, a value of U = 6 eV, suitable for describing the influence of oxygen vacancies on the electronic structure, was also tested.[66] All the crystal structure models are visualized using VESTA.[67]

**2.5. Adsorption energy calculation**. We began by evaluating the overbinding correction ($E_{OB}$) of the molecular O$_2$ and the cohesive energies of the bulk Pt and *anatase*-TiO$_2$ (see Sections S2 and S3 for details). To maintain a reliable comparison between different functionals, all two-dimensional surface lattice parameters were taken from experiment. This setting results in an apparent deviation of the calculated surface energy from the experimental values (Table S3). To ensure calculation accuracy, we conducted convergence tests for the number of layers and cutoff kinetic energy using a surface with adsorbed O$_2$ molecules (Section S1). Section S5 explains the exhaustive calculation of $-E_{ads}$, which is referred to as calculated adsorption enthalpy change ($\Delta H_{calc}$).

To explain the bonding characteristics of O* and O$_2$* on the surfaces, we performed crystal orbital Hamiltonian population (COHP) analysis using LOBSTER.[68–72] The analysis, as presented in Section S7-S9, was conducted to understand why certain functionals perform better or worse for different types of bonding, primarily between the adsorbate and surface.

**2.6. Free energy calculation.** Based on the obtained atomic and molecular O$_2$ adsorption geometries, we exhaustively calculated room-temperature (RT) free energy distribution of ORR intermediates on *anatase*-TiO$_2$(101) using vdW-DF3-opt2 and compared it with that obtained from GGA for the chosen pathway. It is important to note that O$_2$ can be reduced to H$_2$O via a four-electron (4e$^-$) transfer mechanism. Conversely, two-electron (2e$^-$) transfer can produce the undesired $H_2O_2$, which negatively impacts fuel cell performance and durability.[73,74] Here, we focus on 4e⁻ transfer pathway, as its optimization should lead to minimization of the impact of 2e⁻ transfer and improved fuel cell performance. The 4e⁻-transfer associative mechanism proceeds via the following elementary reactions

$$O_2(g) + * \rightarrow O_2^*, \qquad (5)$$

$$O_2^* + (H^+ + e^-) \rightarrow O_2H^*, \qquad (6)$$

$$O_2H^* + (H^+ + e^-) \rightarrow O^* + H_2O(l), \qquad (7)$$

$$O^* + (H^+ + e^-)(H^+ + e^-) \rightarrow OH^* + H_2O(l), \qquad (8)$$

$$OH^* + (H^+ + e^-) \rightarrow 2H_2O(l) + *, \qquad (9)$$

where $O_2^*$, $O_2H^*$, $O^*$, and $OH^*$ are the adsorbed intermediates, while * denotes the free active sites for the intermediate adsorptions.[6,75] The free energy calculation is based on the computational hydrogen electrode model,[1] with the kinetic term neglected. The free energy distribution is calculated at $\Phi = 1.229$ V versus the standard hydrogen electron (SHE).[76] This also corresponds to a zero free energy difference between $O_2(g)$ and $H_2O(l)$.[1] We applied an empirical adjustment to the free energy distribution within GGA to maintain the zero free energy difference.[6] This empirical adjustment is based on NIST-JANAF database.[77] Detailed descriptions are presented in Section S10.

## 3. RESULTS AND DISCUSSION

### 3.1. Reconciling the previous experimental TPD spectra.

Previous works have estimated the $E_{ads}$ on Pt (111) from the Arrhenius fitting of logarithm of desorption rate as the function of inverse temperature at constant coverage. Essentially based on Equation (1), the reported values for molecular ($O_2^*$) adsorption are 0.35 eV (33.5 kJ/mol)[36] or 0.38 eV (37 kJ/mol)[37], and those for atomic ($O^*$) adsorption are 1.78 eV (172 kJ/mol)[36] or 1.97 eV (190 kJ/mol)[37] (as discussed in the Introduction). Furthermore, $E_{ads}$ of monolayer $O_2$ on *anatase*-$TiO_2(101)$ was estimated to be $0.19 \pm 0.02$ eV also based on Equation (1) with an assumed pre-exponential factor of $10^{15}$ s⁻¹.[38] As noted in Introduction section, these estimated $E_{ads}$ values carry large uncertainties, which motivated us to reanalyze these experimental TPD spectra[36,38] using the equilibrium thermodynamics-based methodology.[46]

Figures 1(d-f) present the newly extracted $\rho(E_{ads})$. For the $O_2^*$ desorption on Pt(111), $\rho(E_{ads})$ exhibits a single sharp peak at 0.467 eV, which corresponds to stronger binding than that extracted from previous analysis mentioned above by about 0.1 eV. On the other hand, the $O^*$ desorption exhibits a much broader distribution with the highest $\rho(E_{ads})$ peak at 1.286 eV followed by a shoulder which may consist of multiple peaks; we take the value of 1.486 eV as a representative value for this shoulder. The range of $E_{ads}$ extracted here is about 0.5 eV lower than literature values,[36,37] but what is more important is that this analysis indicates some variety in the $E_{ads}$, which may correspond to different adsorption sites. This is a clear advantage over previous analysis which relies on the Polanyi-Wigner in Equation (1) with a single value for the prefactor and $E_{ads}$. As for $O_2$* desorption on *anatase*-TiO$_2$(101), the extracted peak at 64 K corresponds to the $E_{ads}$ of 0.210 eV, which is close to the previous literature values and corresponds to full monolayer of $O_2$* adsorption.[38]

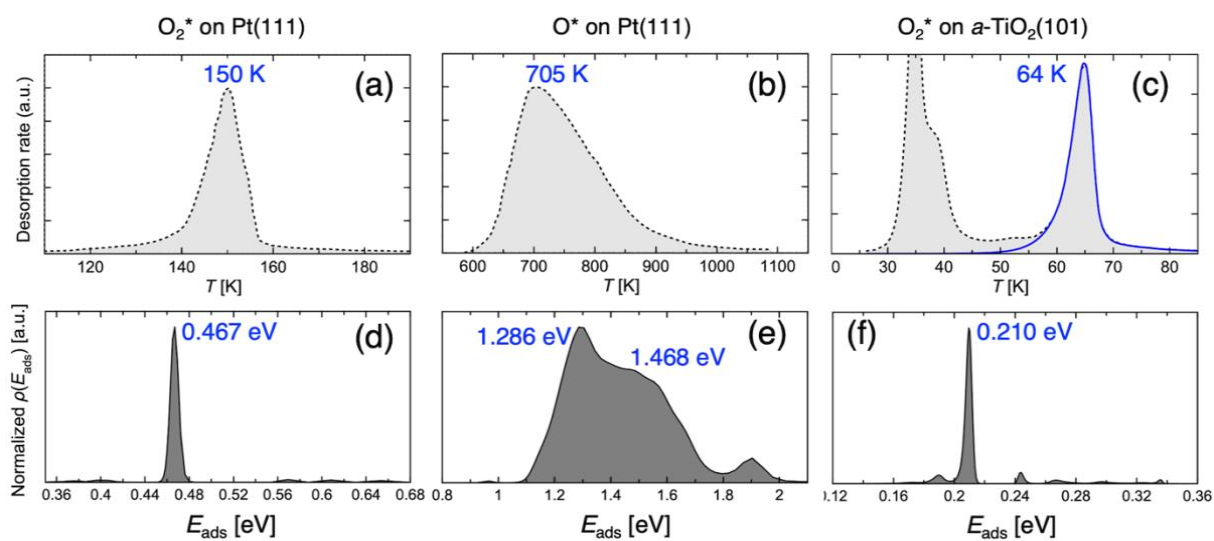

**Figure 1**. (a, b, c) Digitization results (dashed lines) of experimental TPD spectra and (d, e, f) $\rho(E_{ads})$ curves for molecular and atomic desorptions on Pt(111) for the saturation coverage, as well as the molecular desorption on *anatase*-TiO$_2$(101). The blue solid line in (c), corresponding to the saturation coverage, was obtained by assuming exponential decay in the pre-peak region of the dashed line that smoothly connects with the measured TPD spectrum at 3.2 Langmuir.

**3.2. O$_2$ adsorption on Pt(111)**. Figures 2(a) and 2(b) show the structural configurations for $O_2$* and O* adsorption on Pt(111), respectively. For $O_2$* adsorption, we assumed the bridging chemisorbed $O_2$ on Pt(111) as it is known to be the most stable configuration.[14] Notably, the physisorption on Pt(111) acts as a precursor to chemisorption upon heating treatment.[2] Figures 2(c) and 2(d) show the calculated $\Delta H_{calc}$ for $O_2$* and O* adsorption on three and four sites of Pt(111), respectively, across different functionals. For $O_2$* adsorption, we neglected any potential difference in the over-binding correction ($E_{OB}$) related to the O–O bond[21] between the isolated and adsorbed states. This is because no O–O bond dissociation occurs during $O_2$* adsorption. Conversely, $E_{OB}$ was used when calculating $\Delta H_{calc}$ for O* adsorption since the O–O bond dissociates at the adsorbed state (see Section S2). The horizontal dashed lines represent the experimental enthalpy change values obtained from both the literature[36,38] and our $\rho(E_{ads})$, as mentioned in Section 3.1.

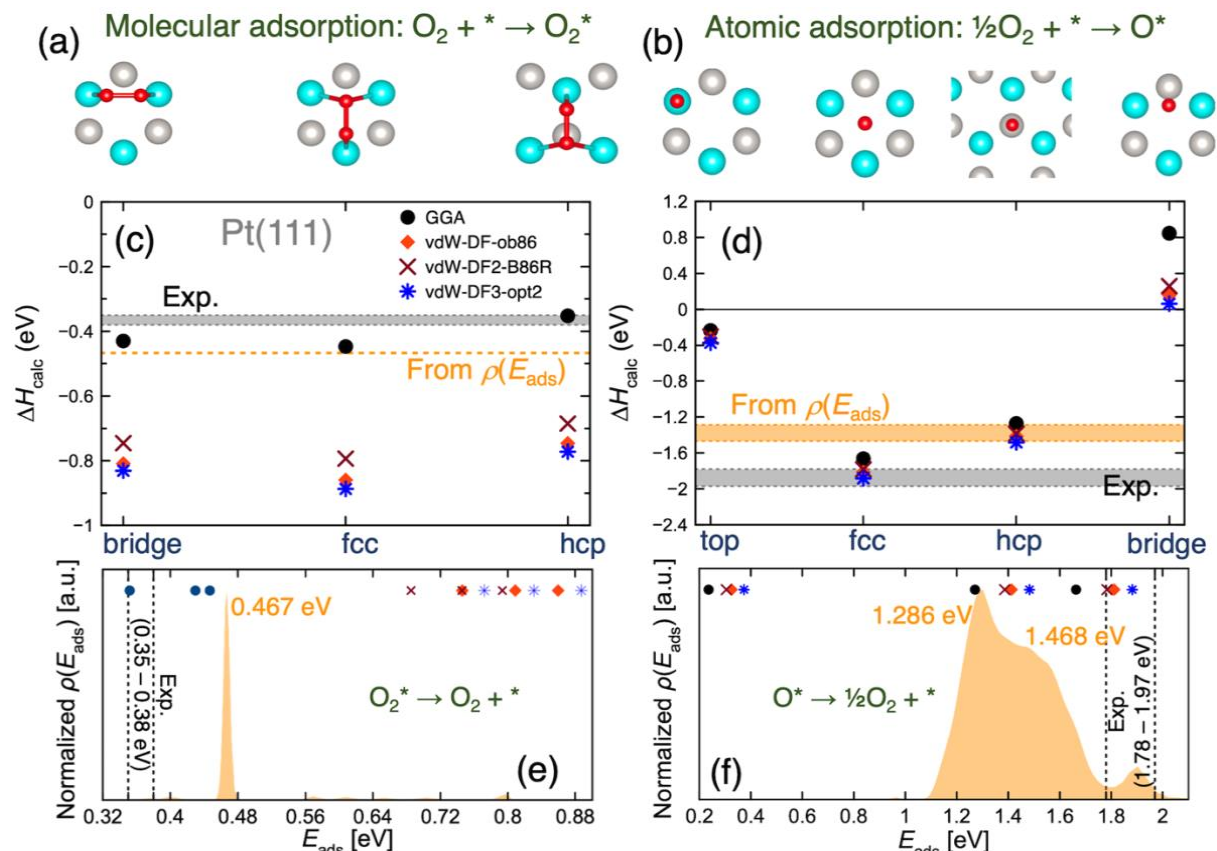

**Figure 2**. (a, b) Top-view molecular and atomic adsorptions of $O_2$ on Pt(111). (c, d) Calculated enthalpy change ($\Delta H_{calc}$) of molecular and atomic adsorptions for different functionals. Horizontal dashed lines denote the experimental enthalpy change estimated by the literature (grey) and from (e, f) $\rho(E_{ads})$ profiles (orange). The vertical dashed lines in (e) and (f) denote the experimental $E_{ads}$ estimated by the literature based on the Arrhenius fitting.

We find that vdW-DF overestimates the adsorption strength of $O_2$*. In contrast, the GGA calculations in this study show good agreement with experimental data. Specifically, our result agrees with another previous research, in which the used lattice parameters were close to experimental values ($a$ = 3.92 Å), and the resulting $E_{ads}$ values were also relatively close to experimental values (0.34–0.45 eV).[16] However, our results contrast with the stronger adsorption reported in several previous GGA calculations with $a \approx 3.99$ Å (see Table S4 for further details). The discrepancy from previous GGA results may be attributed to the differences in the in-plane lattice parameters adopted in this study. Indeed, $E_{ads}$ has been shown to exhibit strong strain dependence for H adsorption on Pt(111),[72] and a similar strain dependence is expected in the present case. Furthermore, there are no significant differences in $E_{ads}$ values between different vdW-DFs, and the $E_{ads}$ values for hcp and fcc sites are very close to each other, as previously confirmed.[2]

For the adsorption of O*, we find that the predicted adsorption from the vdW-DF method is slightly enhanced compared to that of the GGA method, although this enhancement is less pronounced than for $O_2$*. The calculated O* adsorption is strongest at the fcc site and fall within the experimental range. In contrast, values at the hcp site are approximately 0.4 eV smaller and lie outside the experimental range, suggesting the fcc site is energetically favored. However, as Figure 2(f) shows, the values for fcc (hcp) sites are close to smaller (larger) peak of the $\rho(E_{ads})$. Given the significant difference in the peak heights of the $\rho(E_{ads})$, the advantage of hcp sites might be slightly enhanced from an entropic perspective.[46] Considering that the values for top and bridge sites are very small, these sites can be excluded from involvement in the adsorption process.

For all the functionals, the strongest adsorption is found at the fcc site, involving the significant charge transfer from the surface to either $O_2$* or O*.[2] From our analysis, the corresponding ORR pathway may occur at both hcp or fcc sites.[78] Overall, our results indicate that vdW-DFs significantly overestimate the $O_2$* chemisorption $\Delta H_{calc}$ on Pt(111). In contrast, the atomic O* adsorption shows good agreement with the experimental data. Improving the chemisorption strength on Pt(111) would necessitate a more accurate functional. Previous reports have investigated the $E_{ads}$ of oxygen molecules on Pt(111) using various vdW-DFs, generally observing an overestimation trend. However, good agreement with experiments was achieved when using the BEEF-vdW functional with a low vdW contribution, although its performance was not as strong for some other molecules.[31,32] This suggests the inherent difficulty in developing a universally applicable vdW-DF.

**3.3. $O_2$ adsorption on *anatase*-TiO$_2$(101)**. Unlike metal Pt(111) (Figure S5), the electronic structure of *anatase*-TiO$_2$(101) (Figure S8) and its bulk exhibit a large band gap[79] due to their closed-shell character.[80] The non-locality of XC potential is crucial for capturing the interplay between spatial localization and correlation effects in the Kohn-Sham electronic structure.[81] Perturbation theories, such as RPA, are suggested to be suitable to describe the energetics and electronic structure of *anatase*-TiO$_2$-based systems,[82] but are computationally very expensive. To overcome the limitations of RPA, the vdW-DF method offers a practical and suitable alternative for *anatase*-TiO$_2$(101) and its corresponding $O_2$ adsorption.

Figures 3(a) and 3(b) present the calculated $\Delta H_{calc}$ for the molecular and atomic adsorption of $O_2$ on *anatase*-TiO$_2$(101). Both *U* values (*U* = 4 and 6 eV) yield very similar $\Delta H_{calc}$ patterns for almost all the assigned sites (Figure S3), despite the apparent effect of *U* on the electronic structures (Figure S8). The relaxed structures of the molecular adsorption depict $O_2$ physisorption (Figure S10) across all the functionals. GGA underestimates the adsorption strength on *anatase*-$TiO_2$(101) from $\Delta H_{calc}$, a finding consistent with previous results.[14] This underestimation is evident from the lower $E_{ads}$ values estimated from both the literature[38] and our $\rho(E_{ads})$ analysis shown in Figures 3(e) and 3(f). This result suggests that GGA does not properly describe the physisorption. In contrast, the non-local vdW interaction clearly enables $O_2$ binding to the surface, yielding a lower $\Delta H_{calc}$ down to $-0.241$ eV ($E_{ads} = 0.241$ eV). This value is close to the estimated experimental adsorption strength estimated from both the literature and our $\rho(E_{ads})$ analysis. Particularly, $E_{ads}$ for vdW-DF3-opt2 ($E_{ads} = 0.213$ eV) is the closest to the estimated value from $\rho(E_{ads})$ ($E_{ads} = 0.210$ eV), whereas the calculated $E_{ads}$ values for both vdW-DF-ob86 and vdW-DF2-B86R deviate from it. Notably, this $E_{ads}$ corresponds to $O_2$* adsorption at $Ti_{5c}$ site (Figure S12, no. 6). This value generally agrees with results obtained using the empirical rule of the Grimme D3 method and Becke-Johnson decay (0.23 eV).[83] For comparison, the DFT-D2 also provides a corrected $E_{ads}$ of 0.23 eV on rutile $TiO_2$(110), which is stronger than that of GGA (0.08 eV).[84] Additionally, the peak at 34 K might correspond to metastable $O_2$*. Therefore, we suggest that the vdW-DFs, mainly vdW-DF3-opt2, are suitable for describing molecular $O_2$ physisorption on *anatase*-$TiO_2$(101) and possibly for other oxides as well.

Conversely, we found no stable O* adsorption, as indicated by all positive $\Delta H_{calc}$ values in Figures 3(c) and 3(d). For sites 7 and 8 in Figure 3(e), the significant discrepancy observed for vdW-DF3-opt2, compared to the other functionals, stems from distinct relaxed adsorption structures that lead to weak adsorptions. We do not explore this discrepancy further due to limited interest in positive $\Delta H_{calc}$ values. Furthermore, the experimental TPD spectra for *anatase*-$TiO_2$(101) show a peak at around 150 K, which was suggested to be from the adsorption at step sites.[38] Another possibility is that this peak might correspond to O* adsorption in the presence of defects. This might be similar with the case of negative $\Delta H_{calc}$ of atomic adsorption on reduced rutile $TiO_2$(110) with one bridging-oxygen vacancy.[85] Since oxygen vacancy sites are expected to be resistant to $H_2O$ poisoning,[7] future TPD could potentially utilize the current methodology for defective $TiO_2$ surfaces without explicitly modeling pre-adsorbed $H_2O$ layers.

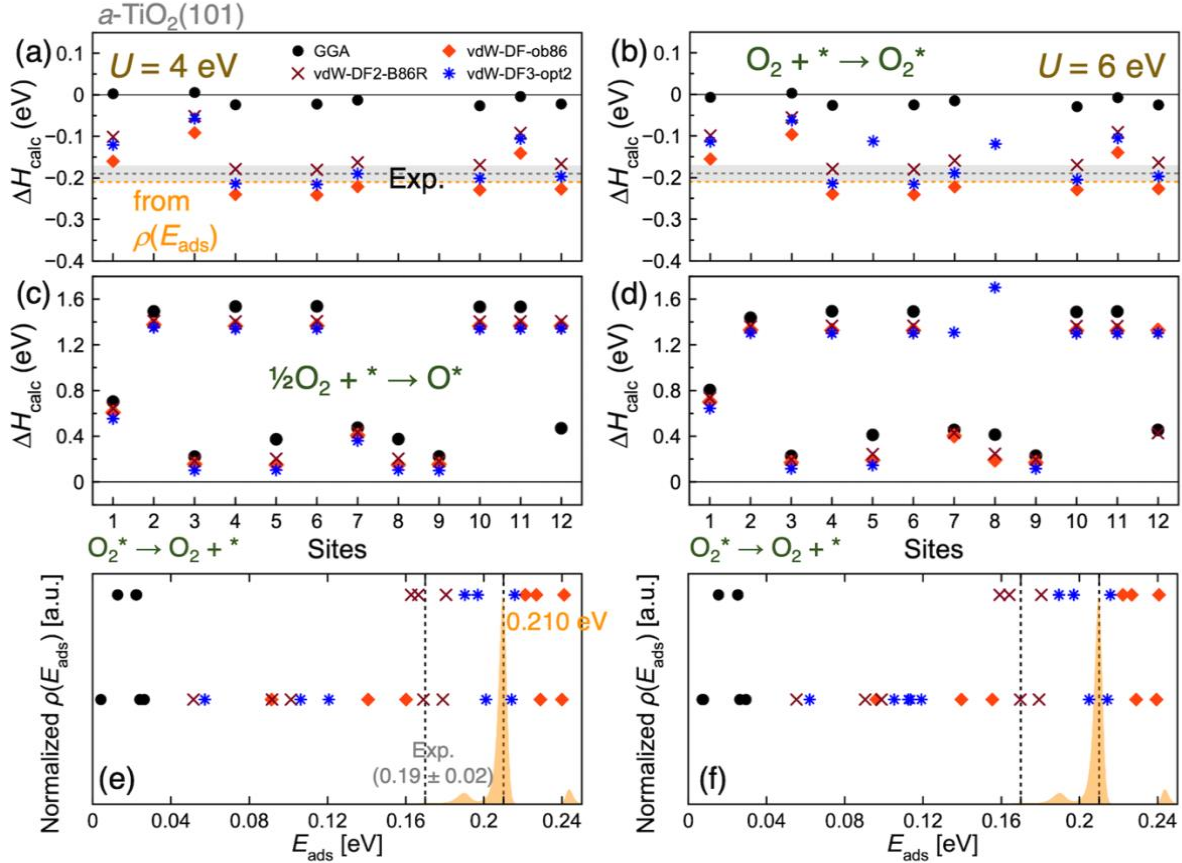

**Figure 3**. Calculated enthalpy change ($\Delta H_{calc}$) of (a, b) molecular and (c, d) atomic adsorptions on *anatase*-TiO$_2$(101) for $U$ = 4 (left) and 6 eV (right). Horizontal dashed lines denote the experimental enthalpy change estimated by the literature (grey) and from (e, f) $\rho(E_{ads})$ profiles (orange). The vertical dashed lines in (e) and (f) denote the experimental $E_{ads}$ estimated by the literature based on Equation (1).

The driving force of the physisorption is most likely induced by dispersion forces, which essentially cause no charge transfer between O$_2$* and the surface. Physisorption is also typically observed on inert Au surfaces due to the fully occupied d orbitals, where TPD peaks were found at low temperatures of 37 and 55 K.[2] Despite the presence of a less coordinated Ti$_{5c}$ site, *anatase*-TiO$_2$(101) exhibits an inert state. Furthermore, the presence of surface host oxygen atoms is thought to significantly suppress charge transfer to O$_2$* and to effectively act as a repulsive region for O.*[38,86] If the surface is defective, however, this behavior may change, potentially allowing atomic adsorption to occur.[85]

**3.4. Electronic and magnetic properties of adsorption systems**. In this subsection, we examine the bonding and magnetic properties of O* and O$_2$* adsorbed on Pt(111) and *anatase*-TiO$_2$(101). Our analysis utilizes total density of states (DOS), projected DOS (PDOS), projected COHP (pCOHP), and integrated pCOHP (IpCOHP), presented in the Supporting Information. For both the clean surfaces and surface-adsorbate systems, the quantities obtained using all functionals are very similar to each other. Hence, we focus here on IpCOHP obtained for the most stable adsorbate structure at the Fermi level ($E_F$) and comment on its functional-dependence behavior. It is worth noting that, as shown in Figure S4, isolated $O_2$ exhibits the spin-polarized triplet state ($^3O_2$), which is more stable than its singlet counterpart ($^1O_2$).[18]

Figures S5, S6, and S7 confirm that the metallic behavior of Pt(111) primarily stems from surface Pt 5d states at the Fermi level ($E_F$). The contributions from O 2p states of $O_2$* and O* remain minimal and non-magnetized. For $O_2$* on Pt(111), O–O bond of $O_2$* structure is stronger than O-surface Pt bonds, but weaker than that of the isolated $O_2$. Furthermore, the relatively strong O* adsorption likely correlates with the predominant covalent bond formed between the O atom and surface Pt, as shown by pCOHP in Figure S7. These results are in contrast with Au(111) with the fully occupied d orbital, which supports the molecular physisorption only.[2]

For $O_2$* on *anatase*-$TiO_2$(101), Figure S8 and S9 show the minor effect of the Hubbard $U$ correction, leading us to focus on the bonding characteristics for $U$ = 4 eV. Unlike the case for Pt(111), Figure S10 shows that physisorption maintains the spin-polarized adsorbed triplet state of $^3O_2$*. Here, the O–O bond also remains stronger than the O-surface Ti bond based on IpCOHP. Furthermore, Figure S11 indicates a zero magnetic moment of O*. Despite the finite IpCOHP of O-surface Ti bond, the absence of stable atomic adsorption could be correlated with the inert-like behavior of *anatase*-$TiO_2$(101), which is similar to that of inert Au(111), which also avoids the atomic adsorption.[2]

**3.5. Free energy distribution of ORR on *anatase*-$TiO_2$(101)**. Building upon the calculated adsorption properties of $O_2$ on *anatase*-$TiO_2$(101), we further evaluate the ORR by investigating vdW-DF contributions. Using vdW-DF3-opt2 for $U$ = 4 eV, we calculated the room temperature (RT) 4e⁻-transfer free energy distribution of ORR at $\Phi$ = 1.229 V, as presented in Figure 4(a). Based on the preferable ORR pathway (discussed in Section 3.6), this free energy diagram is compared with that of GGA (at the same $U$ value) as a reference, using the same initial sites for intermediates, as depicted in Figure 4(b). This comparison aims to further evaluate the contribution of non-local interactions between the intermediates and the surface. The associative mechanism is assumed to cover all relevant intermediates and the surface, as indicated in Equations (5)–(9). Notably, the free energy of each step is independent of that of the previous and subsequent intermediates (see Section S10). Figures S12-S15 present all the relaxed structures of intermediates with each of their $\Delta G$ values at $\Phi$ = 0 V for vdW-DF3-opt2.

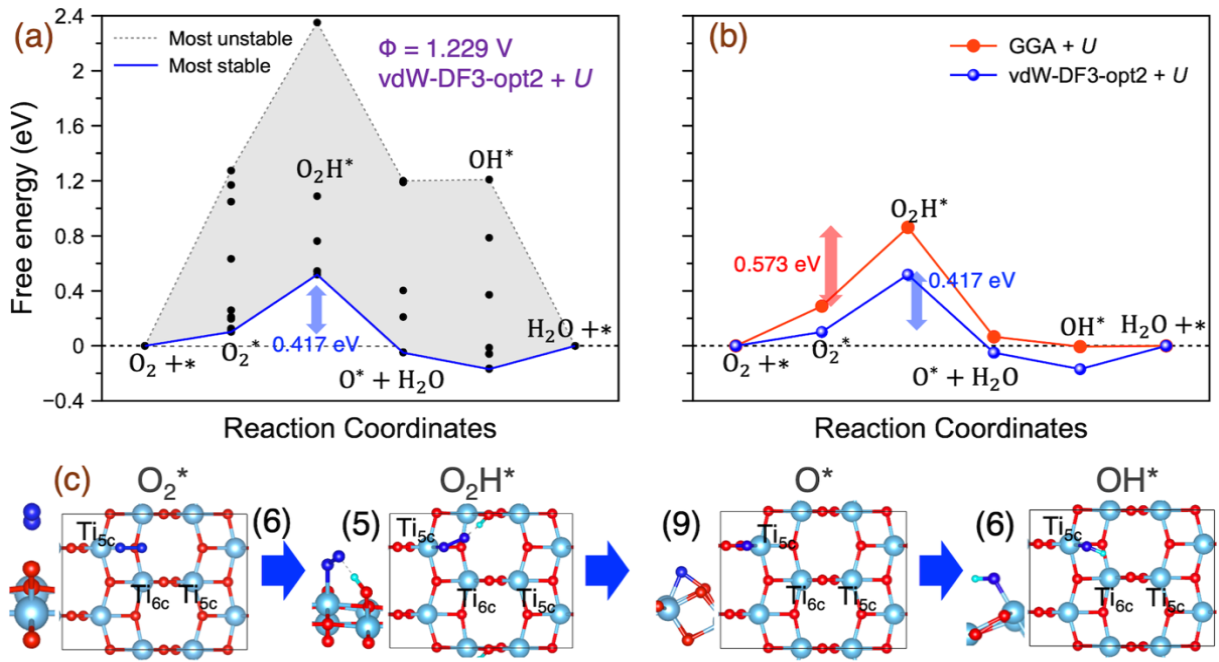

**Figure 4**. (a) Calculated free energy distribution of oxygen reduction reaction on *anatase*-TiO$_2$(101) at $\Phi$ = 1.229 V using vdW-DF3-opt2 + $U$. Solid blue and dashed grey lines denote the free energy pathways for the most stable and unstable intermediate structures, respectively. (b) Free energy diagram of the ORR for the chosen pathway (blue) compared with that of GGA + $U$ (red). (c) The relaxed most-stable intermediates are modelled for vdW-DF3-opt2 + $U$. The numbers are from the initial site assignment (see Figure S3).

**3.6. Preferable ORR pathway**. The free energy distribution provides numerous ORR pathways. To elucidate the preferable reaction pathway, we elaborate on the relative free energy of the most stable intermediate structures for vdW-DF3-opt2, as shown by the solid blue line in Figure 4(c). The ORR is initiated by the endothermic O$_2$ physisorption, with $\Delta G_1$(O$_2$*) of 0.101 eV (Figure 4(a)), where one adsorbed O atom is bound to the Ti$_{5c}$ site. In contrast, GGA shows the less stable physisorption with a $\Delta G_1$(O$_2$*) of 0.289 eV (Figure 4(b)), influenced by the underestimated $E_{ads}$ (Figure 3(a)). The first electron (e$^-$) transfer involves O$_2$H* formation, with an endothermic $\Delta G_2$(O$_2$H*) of 0.417 eV. This represents the maximum free energy change ($\Delta G_{max}$), possibly corresponding to the rate-determining step (RDS). This $\Delta G_{max}$ is lower than that of Pt(111), i.e., 0.45 eV for OH* formation.[1] The formation involves the transition from O$_2$* physisorption to O$_2$H* end-on chemisorption, where the same adsorbed O atom remains bound with Ti$_{5c}$. However, less stable O$_2$H* with $\Delta G_2$(O$_2$H*) of 0.573 eV is obtained for GGA. This free energy difference between both functionals is attributed to the weaker interaction between O$_2$H* and the surface resulting from the absence of non-local correction. The second e$^-$ transfer is the formation of O* and the first H$_2$O, with an exothermic $\Delta G_3$(O*) of –0.568 eV. This e$^-$ transfer leads to the adsorbed O atom being bound to both Ti$_{5c}$ and one surface O atom. Furthermore, GGA shows a lower $\Delta G_3(O^*)$ of $-0.796$ eV. The third e$^-$ transfer is OH* formation with the exothermic $\Delta G_4(OH^*)$ of $-0.120$ and $-0.072$ eV for vdW-DF3-opt2 and GGA, respectively. This indicates that OH* formation is the most stable intermediate within the preferable pathway. During this transfer, H$^+$ replaces the O*–surface O bond to form H–O–Ti$_{5c}$ bonds. The final e$^-$ transfer involves the second H$_2$O formation with $\Delta G_5(H_2O)$ of 0.170 and $-0.006$ eV using vdW-DF3-opt2 and GGA, respectively. This result contrasts with the final transfer on the pristine $t$-ZrO$_2$(101) as the RDS due to strong OH* adsorption relative to the free energy of H$_2$O($l$).[6,75]

Finally, we suggest that Ti$_{5c}$ is the active site for attaining the preferable ORR pathway. Our results demonstrate that lower-coordinated metal centers, such as Ti in this case, is more reactive toward O$_2$ and oxygen-containing species, including ORR intermediates.[2] Furthermore, given the less stable O$_2$* and O$_2$H* adsorptions within GGA, the ORR is more likely to occur and is described more accurately by vdW-DF3-opt2, especially considering the simulated TPD spectra of O$_2$ physisorption. Considering that no defects were explicitly involved in our simulation, we propose that pristine *anatase*-TiO$_2$(101) could serve as an optimal catalyst. This is due to the intermediate binding strengths of the reactant (O$_2$*) and the intermediates obtained with vdW-DF3-opt2, which align with the Sabatier principle.[87]

## 4. CONCLUSION

We have investigated the reliability of $E_{ads}$ of O$_2$ on Pt(111) and *anatase*-TiO$_2$(101), as calculated using vdW-corrected GGA functionals. Our approach involved comparing the calculated $E_{ads}$ values with simulated $\rho(E_{ads})$ derived from experimental TPD spectra. For *anatase*-TiO$_2$(101), the experimental TPD spectrum can be quantitatively elucidated by the physisorbed molecular state O$_2$*. Here, the non-local vdW correction plays a crucial role in accurately describing the adsorption. In contrast, for Pt(111), the $E_{ads}$ of O$_2$* is considerably overestimated by vdW-corrected functionals compared to experimental values, while the $E_{ads}$ of O* shows reasonable agreement with the experimental data. t is important to note that the $E_{ads}$ accuracy of O$_2$* has not been thoroughly assessed in the literature, making their verification against experimental data particularly valuable. Our findings suggest that thermodynamic properties of reactions initiated by O$_2$ adsorption, such as the ORR via the associative pathway on *anatase*-TiO$_2$(101) can be experimentally validated. However, accurate modeling of O$_2$ adsorption on Pt(111) still requires further refinement of the exchange–correlation functional.

- **ACKNOWLEDGMENTS**

This research was supported by the New Energy and Industrial Technology Development Organization (NEDO) and the Digital Transformation Initiative for Green Energy Materials (DX-GEM) projects. All calculations were performed at the Supercomputer Center of the Institute for Solid State Physics (ISSP), The University of Tokyo, Japan.

**Notes**

There are no conflicts of interest to declare.

**Supporting Information Available**

# Supporting Information: Benchmarking of Oxygen Adsorption using TPD Spectroscopy for Accurate DFT Prediction of ORR on Anatase Titanium Dioxide (101)

## S1. Calculation details

This work employs four exchange-correlation functional types, i.e., Perdew-Burke-Ernzerhof type generalized gradient approximation (GGA)[1] and three van der Waals density functionals (vdW-DFs), i.e., vdW-DF-ob86,[2] vdW-DF2-B86R,[3,4] and vdW-DF3-opt2.[5] The projector augmented wave (PAW) method[6] for pseudopotentials is used.

The isolated $O_2$ structure was built with the experimental bond length ($d_{O-O}$) of 1.208 Å.[7] The corresponding energy calculation employed Γ $k$-point ($k = 0$) within the Monkhorst-Pack scheme.[8] Martyna-Tuckerman correction is used for the isolated $O_2$,[9] located at the center of a large cubic unit cell with a unit dimension of 20 × 20 × 20 Å$^3$. Platinum (Pt) (111) was built from the experimental crystal structure of bulk cubic Pt ($a$ = 3.9231 Å, space group: $Fm\bar{3}m$, no. 225).[10] For the cubic system, a k-point mesh of $k \times k \times k$ was optimized using a convergence test, leading to the mesh of 7 × 7 × 7 as depicted in Figure S1(a). Anatase titania (*anatase*-TiO$_2$) (101) was built from the experimental structure of bulk *anatase*-TiO$_2$ ($a$ = 3.7845 Å, $c$ = 9.5143 Å, $z_O$ = 0.20806, space group: $I4_1/amd$, no. 141).[11] For the tetragonal system, a k-point mesh of $q \times q \times k$ was optimized using a convergence test previously used for layered oxychalcogenides.[12] From the lattice parameters, the magnitudes of reciprocal vectors (1/$a$, 1/$a$, 1/$c$) provide equal densities where (1/$a$)/$q$ = (1/$c$)/$k$ or $q/k = c/a$. For the bulk *anatase*-TiO$_2$, the value of $c/a$ is ~2.51, resulting in a rounded-down value of $q/k \approx 2$. Then, we tested for $k$ of 2, 3, 4, and 5, as we expect that the value of $k = 1$ is too small for the unit cell of bulk *anatase*-TiO$_2$. A $k$-points mesh of 8 × 8 × 4 was chosen as shown in Figure S1(e), which similar to that of the tetragonal ZrO$_2$.[13] In Figure S1(b) and S1(f), convergence tests of cut-off kinetic energies were performed for the bulk Pt and *anatase*-TiO$_2$, from which the cut-off kinetic energies of 100 Ry (~1360 eV) and 80 Ry (~1090 eV), respectively, were chosen to determine the total energy of each system. The valence electron configurations of elements used in this work are H: 1s$^1$, O: 2s$^2$ 2p$^4$, Pt: 5d$^9$ 6s$^1$ 6p$^0$, and Ti: 3s$^2$ 3p$^6$ 3d$^2$ 4s$^2$. The Hubbard corrections ($U$) of 4.0 and 6.0 eV are used for Ti 3d orbital.[14–16] The structural optimization was performed using the Broyden–Fletcher–Goldfarb–Shanno (BFGS) method[17–21] with a force threshold of 0.02 eV/Å, while the energy calculation used a threshold energy of 10$^{-7}$ a.u. (~2.7 × 10$^{-6}$ eV).

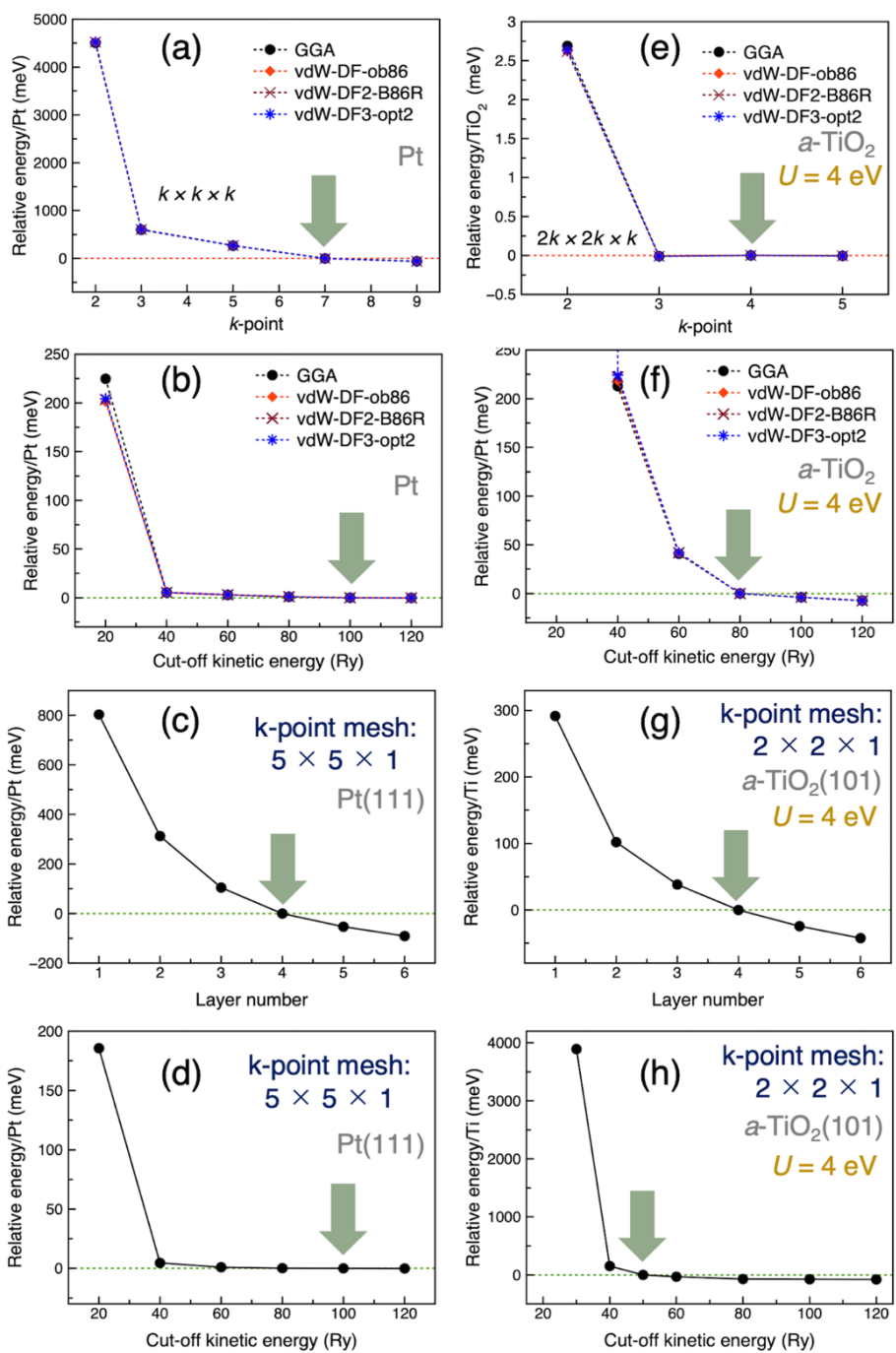

**Figure S1**. Convergence tests of k-point mesh and cut-off kinetic energy of (a, b) bulk Pt and (e, f) *anatase*-TiO$_2$. Convergence tests of layer number and cut-off kinetic energy of (c, d) Pt(111) and (g, h) *anatase*-TiO$_2$(101) are also presented.

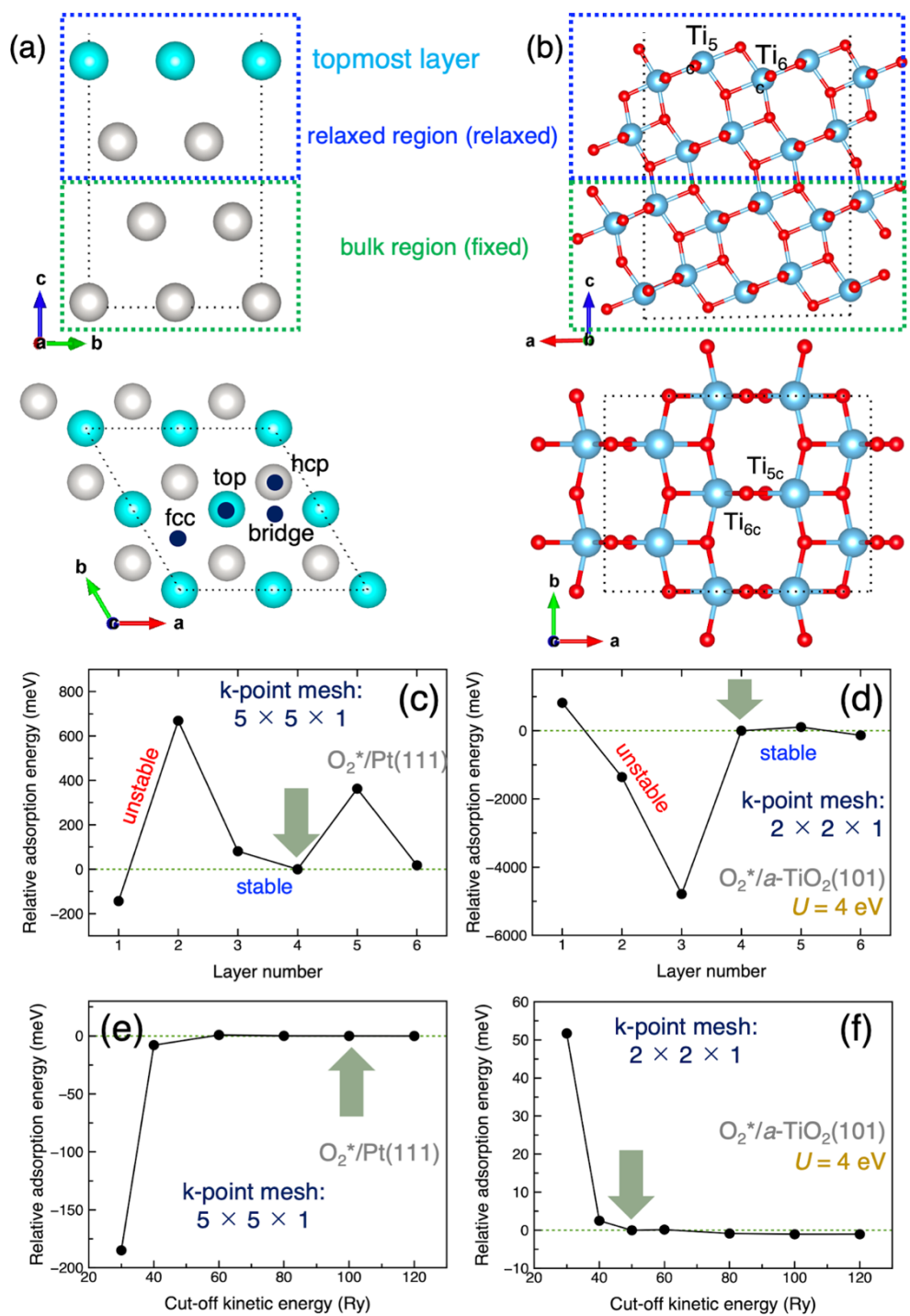

**Figure S2**. Crystal models of (a) Pt(111) and (b) *anatase*-TiO$_2$(101) from side and top views. Grey, light blue, and red colors denote Pt, Ti, and O atoms. Dashed lines denote unit cell boundaries. Four adsorption sites of Pt(111) are assigned: fcc, top, hcp, and bridge sites. The five- and six-coordinated Ti atoms (Ti$_{5c}$, Ti$_{6c}$) at the topmost layer of *anatase*-TiO$_2$(101) are also assigned. Convergence tests of layer number and cut-off kinetic energy of (c, e) Pt(111) and (d, f) *anatase*-TiO$_2$(101) are presented. The green arrows denote each chosen layer number and cutoff kinetic energy.

Figures S2(a) and S2(b) show the structural model of Pt(111) and *anatase*-TiO$_2$(101) based on each bulk counterpart. VESTA[22] was used for all crystal model visualizations in this work. Pt(111) was built as a hexagonal-type 2 × 2 supercell slab with a 2D unit dimension of 5.5481 × 5.5481 Å$^2$, corresponding to Pt(111) surface density of 1.5 × 10$^{15}$ mol/cm$^2$, as also indicated by the previous experimental work.[23] On the other hand, the *anatase*-TiO$_2$(101) was built as 96-atoms of a monoclinic-type 2 × 1 slab supercell with the 2D dimension of 10.2394 × 7.5690 Å$^2$ with $β$ (∠ac) of 91.0427°. To avoid any interaction between periodic perpendicular slab structures, a vacuum layer of ~12 Å is used for each slab structure. The reduced k-point meshes of 5 × 5 × 1 and 2 × 2 × 1 are used for Pt(111) and *anatase*-TiO$_2$(101), respectively, to reduce computational cost. While the threshold force for structural optimization was kept the same as that for the bulk system, reduced threshold energy of 10$^{-6}$ a.u. (~2.7 × 10$^{-5}$ eV) was used for the corresponding energy calculation. A convergence test of layer number is also presented in Figures S1(c) and S1(g), confirming that the four-layer model for both surfaces is sufficient. Figure S2(c) and S2(d) present the convergence tests of layer number dependence of molecular adsorption energy ($E_{ads}$) of O$_2$ on Pt(111) and *anatase*-TiO$_2$(101) ($U$ = 4 eV), respectively, confirming that the layer number of 4 also shows stable $E_{ads}$, while unstable $E_{ads}$ is found for the fewer than four layers. This layer number leads to 16-atoms Pt(111) and 96-atoms *anatase*-TiO$_2$(101) supercells. $E_{ads}$ is defined as the negative of $\Delta H_{calc}$ in Equation (S5). A convergence test of cut-off energy is presented in Figure S1(d) and S1(h), confirming that cut-off kinetic energies of 100 and 50 Ry are sufficient for Pt(111) and *anatase*-TiO$_2$(101) ($U$ = 4 eV), respectively. These chosen values are also supported by the convergence test of cut-off energy dependence of molecular $E_{ads}$ of O$_2$ in Figures 2(e) and 2(f). The different functionals provide no significant energy differences, thus, the convergence test was performed within GGA.

## S2. Overbinding correction of O$_2$

The binding energy ($E_B$) of the isolated O$_2$ is calculated as the energy difference to form O$_2$(g) ($E_{O_2}$) from two O atoms ($E_O$) formulated as 2O → O$_2$(g). $E_B$ is expressed as

$$E_B = E_{O_2} - 2E_O \text{ (in eV)} \tag{S1}$$

The experimental binding energy ($E_B^{exp}$) of 5.116 (NIST-JANAF)[24,25] are used to calculate overbinding correction ($E_{OB}$) using

$$E_{OB} = E_B + E_{ZPE} - E_B^{exp} \text{ (in eV)} \tag{S2}$$

where $E_{ZPE}$ is the zero-point energy.[24] $E_{ZPE}$ values of O$_2$ are calculated for each functional, which is close to the experimental value of 99.1 meV.[26] We assume that $E_{ZPE}$ is inherently included in $E_\text{B}^\text{exp}$. As for the adsorbed O$_2$ (O$_2$*, see Section S5), $E_\text{ZPE}$ is adopted from the previous research, i.e., 111 and 74 meV for molecular and atomic adsorptions, respectively,[27] leading to the zero-point energy change ($\Delta E_\text{ZPE}$) to adsorb O$_2$. Table S1 presents the calculated O–O bond length ($d_\text{O–O}$), $E_\text{ZPE}$, $E_\text{B} + E_\text{ZPE}$, and $E_\text{OB}$ of the isolated O$_2$ structure for different functionals.

**Table S1.** Calculated $d_\text{O–O}$, $E_\text{ZPE}$, $E_\text{B} + E_\text{ZPE}$, and $E_\text{OB}$.

| Functional | $d_\text{O–O}$ [Å] | $E_\text{ZPE}$ [meV] | $E_\text{B} + E_\text{ZPE}$ [eV] | $E_\text{OB}$ [eV] |
|---|---|---|---|---|
| GGA | 1.2288 | 95.9 | -6.559 | -1.444 |
| vdW-DF-ob86 | 1.2302 | 95.1 | -6.351 | -1.236 |
| vdW-DF2-B86R | 1.2303 | 95.1 | -6.368 | -1.253 |
| vdW-DF3-opt2 | 1.2292 | 94.9 | -6.444 | -1.328 |

**Table S2.** Cohesive energy ($E_\text{coh}$) and its difference $|\Delta E_\text{coh}|$ of Pt and *anatase*-TiO$_2$(101)

| Functional | $E_\text{coh}$ ($|\Delta E_\text{coh}|$) | | | |
|---|---|---|---|---|
| | Pt [in eV/Pt] | *anatase*-TiO$_2$ [in eV/Ti] | | |
| | | $U = 0$ eV | $U = 4$ eV | $U = 6$ eV |
| GGA | 5.73 (0.11) | 21.31 (1.47) | 17.57 (2.27) | 16.00 (3.84) |
| vdW-DF-ob86 | 6.30 (0.46) | 17.57 (2.27) | 17.79 (2.05) | 16.30 (3.54) |
| vdW-DF2-B86R | 6.15 (0.31) | 21.56 (1.73) | 17.96 (1.88) | 16.38 (3.46) |
| vdW-DF3-opt2 | 6.26 (0.42) | 17.79 (2.05) | 18.30 (1.54) | 16.72 (3.11) |

## S3. Cohesive energy

To confirm the structural stability of the bulk systems, Table S2 presents cohesive energies ($E_\text{coh}$) of both bulk Pt and *anatase*-TiO$_2$ for different functionals, compared with the experimental values of $E_\text{coh-exp}$ = 5.84[28] and 19.83 eV,[29] respectively. For the bulk *anatase*-TiO$_2$, the results obtained using different values of U are also presented. The quantity $|\Delta E_\text{coh}|$ in parentheses indicates the absolute deviation between the calculated and experimental $E_\text{coh}$, which is expressed as

$$E_{coh} = (\sum_i n_i E_i - E_\text{bulk})/F \text{ (in eV)} \tag{S3}$$

Where $E_\text{bulk}$ is the total energy of the bulk system, $n_i$ and $E_i$ are the number and the isolated-state energy of *i*-th atom, while $F$ is either the number of formula units or the number of atoms per unit cell. For the bulk Pt, $n_1$ and $F$ are equal to 1. For the bulk *anatase*-TiO$_2$, depicting Ti$_2$O$_4$ for the space group of $I4_1/amd$, the value of $F = 2$ is used as the number of formula units.

## S4. Surface energy

Structural stability of the surfaces is quantified by the surface energy ($\gamma_{surf}$) expressed as

$$\gamma_{surf} = [E_{surf} - E_{bulk}(N_{surf}/N_{bulk})]/2A \text{ [in J m}^{-2}] \tag{S4}$$

where $E_{bulk}$ and $E_{surf}$ are the total energies of the bulk and the surface structures, respectively. $N_{bulk}$ and $N_{surf}$ are the total number of atoms in the bulk and surface structures, respectively. For Pt(111), $N_{bulk}$ and $N_{surf}$ are 1 and 16, respectively. On the other hand, $N_{bulk}$ and $N_{surf}$ for *anatase*-TiO$_2$(101) are 6 and 96, respectively. $A$ is the surface area. Table S3 presents $\gamma_{surf}$ of both surfaces. The experimental $\gamma_{surf}$ of 2.49 J/m$^2$ derived from the liquid surface tension measurement[30] is used as the reference for Pt(111). On the other hand, the experimental $\gamma_{surf}$ of 0.4 ± 0.1 J/m$^2$ from the high-temperature oxide melt drop solution calorimetry[31] is used as the reference for *anatase*-TiO$_2$(101). The quantity $|\Delta\gamma_{surf}|$ [in %] in the parentheses is the absolute value of the difference between the calculated and experimental $\gamma_{surf}$ values divided by the experimental value. Note that the surface unit cell dimensions are directly derived from the experimental lattice parameters of the bulk phases, which may contribute to the observed large values of $\gamma_{surf}$.

**Table S3.** Surface energy ($\gamma_{surf}$) and its difference $|\Delta\gamma_{surf}|$ of Pt(111) and *anatase*-TiO$_2$(101)

| Functional | $\gamma_{surf}$ ($|\Delta\gamma_{surf}|$) [in J/m$^2$] | | | |
|---|---|---|---|---|
| | Pt(111) | *anatase*-TiO$_2$(101) | | |
| | | $U = 0$ eV | $U = 4$ eV | $U = 6$ eV |
| GGA | 1.19 (1.30) | 1.17 (0.77) | 1.17 (0.77) | 1.06 (0.66) |
| vdW-DF-ob86 | 1.72 (0.77) | 1.05 (0.65) | 1.05 (0.65) | 1.42 (1.02) |
| vdW-DF2-B86R | 1.66 (0.83) | 1.36 (0.96) | 1.36 (0.96) | 1.36 (0.96) |
| vdW-DF3-opt2 | 1.68 (0.81) | 1.40 (1.00) | 1.40 (1.00) | 1.39 (0.99) |

## S5. Adsorption enthalpy changes of molecular (O$_2$*) and atomic (O*) adsorptions

For Pt(111), we focus on the adsorption on three and four sites for the molecular and atomic adsorptions, leading to adsorbed O$_2$* (bridge, fcc, hcp) and O* (top, fcc, hcp, bridge), respectively [see Figure S2a]. The term * denotes the adsorption site. As for *anatase*-TiO$_2$(101), Figure S3 shows the exhaustive sampling of initial adsorption sites of O$_2$* and O* on *anatase*-TiO$_2$(101). As the initial adsorption position was varied, the single orientation of O$_2$ is used, as shown in Supporting Information of our previous work,[32] to reduce a possible massive number of calculations from multiple orientations. The corresponding enthalpy change ($\Delta H_{calc}$) of $O_2^*$ and $O^*$ is defined by[33]

$$\Delta H_{calc} = E_{total} - E_{surf} - E_{adsorbate} + \Delta E_{ZPE} \text{ (in eV)} \tag{S5}$$

where $E_{total}$, $E_{surf}$, and $E_{adsorbate}$ are the total energy of the surface with adsorbate, the surface without adsorbate, and an isolated molecule, respectively. $\Delta E_{ZPE}$ is the zero-point energy change (see Table S1). $\Delta H_{calc}$ for $O^*$ and $O_2^*$ adsorptions correspond to the reactions $O + * \rightarrow O^*$ and $O_2 + * \rightarrow O_2^*$, respectively. Table S4 shows $\Delta H$ across the literature for both experiments from temperature-programmed desorption (TPD) spectroscopy and calculations.

**Table S4**. Adsorption energies ($E_{ads}$) of $O_2$ adsorption on Pt(111) and *anatase*-TiO$_2$(101) from experiments and calculations

| Method | Bulk lattice constant ($a$) | Adsorption enthalpy change ($\Delta H$) [eV] | |
|---|---|---|---|
| | | Molecular ($O_2^*$) | Atomic ($O^*$) |
| Surface: Pt(111) | | | |
| TPD[34] | – | 0.35 (33.5 kJ/mol) | 1.78 (172 kJ/mol) |
| TPD[35] | – | 0.38 (37 kJ/mol) | 1.97 (190 kJ/mol) |
| PBE-GGA[36] | 4 Å | 0.46 (hcp) <br> 0.64 (bridge) <br> 0.65 (fcc) | 1.08 (fcc) |
| PBE-GGA[37] | 3.99 Å | – | 0.99 (hcp) <br> 1.65 (fcc) |
| PBE-GGA[38] | 3.985 Å | – | ~1.3 (fcc) |
| PBE-GGA[39] | 3.99 Å | 0.45 (hcp) <br> 0.63 (bridge) <br> 0.65 (fcc) | – |
| PBE-GGA[40] | 3.923 Å (experiment) | 0.072 (physisorption) <br> 0.34 (hcp) <br> 0.42 (fcc) <br> 0.45 (bridge) | – |
| PBE-GGA[41] | 3.92 Å (experiment) | 0.02 (physisorption) <br> 0.39 (end-on chemisorption) <br> 0.69 (bridging chemisorption) | – |
| PBE-GGA[42] | $a$ = 3.983 Å | 0.60 (hcp, non-interacting) <br> 0.66 (hcp, interacting) <br> 0.74 (bridge, non-interacting) <br> 0.76 (brige, interacting) <br> 0.79 (fcc, non-interacting) <br> 0.74 (fcc, interacting) | 2.0 ($O_2^*$-bridge to $O^*$-hcp + fcc) <br> 2.2 ($O_2^*$-fcc to $O^*$-2 fcc), <br> 2.4 ($O_2^*$-bridge to $O^*$-2 fcc) |
| Surface: *anatase*-TiO$_2$(101) | | | |
| TPD[43] | | 0.19 ± 0.02 eV | – |

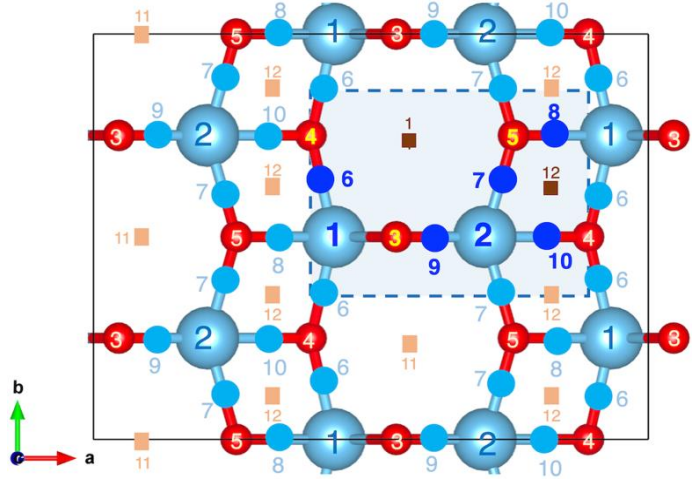

**Figure S3**. Assignment of initial adsorption sites of $O_2^*$ and $O^*$ on *anatase*-TiO$_2$(101). Light blue and red colors denote Ti and O atoms, respectively. Dark-colored dots denote the initial intermediate positions within the topmost-layer primitive two-dimensional (2D) cell, while light-colored dots denote the corresponding symmetrical positions. Notably, Ti$_{6c}$ and Ti$_{5c}$ are denoted at sites 1 and 2, respectively.

## S6. Temperature-programmed desorption spectroscopy

The TPD[44] can be used for obtaining information about $E_{\text{ads}}$ of molecules based on the Polanyi–Wigner equation,[45,46] where the desorption rate is expressed as

$$-\frac{d\theta}{dt} = \theta^n \nu_0 e^{-\Delta E/k_B T} \tag{S6}$$

where $k_B$ is Boltzmann constant, $t$ is the time, $\theta$ is the coverage, $n$ is the desorption order, $\nu_0$ is the pre-exponential factor, $T$ is the temperature, $\Delta E$ is the desorption barrier or the energy difference between the gas and adsorbed phases. The equilibrium thermodynamics is used in the present work, where both adsorption and desorption are independent of each other with the same rate but in opposite directions. The desorption rate, which is equal to the negative adsorption rate, is expressed as

$$-n_a \frac{d\theta}{dt} = s N_d \frac{p}{\sqrt{2\pi m k_B T}} \tag{S7}$$

where $p$ is the pressure, $m$ is the gas molecule mass, $n_a$ is the density of adsorption site (per area), factor $N_d$ is 1 or 2 for molecular or atomic adsorptions, respectively, and $s$ is the sticking coefficient at the desorption temperature, depending on the kinetic, vibrational, or rotational energy of the molecules. In the equilibrium state, the chemical potential of the adsorbed ($\mu_a$) and gas phases ($\mu_g$) are equal, expressed as

$$N_d \mu_a = \mu_g = \mu_{g0} + k_B T \ln \frac{p}{p_0} \tag{S8}$$

where $\mu_{g0}$ is the chemical potential at the gas phase at reference pressure $p_0$ ($p_0 = 10^5$ Pa).

For a single adsorption, the adsorbate can be assumed as ideal 3D, 2D, or lattice gases, determining its entropy. However, the adsorption does not realistically correspond to several single adsorption, each of which is treated as an independent adsorption. Instead, $E_{ads}$ (without the vibrational, rotational, and translational contributions) should be treated as a distribution of adsorption energies, with which the TPD spectrum is not simply from TPD spectra superposition of multiple $E_{ads}$. In equilibrium, each site of a lattice gas can be occupied by one or zero molecules assumed. This condition can be described by the Fermi-Dirac statistics, which is irrespective of spins. Furthermore, the adsorption sites are the microstates in Fermi-Dirac thermodynamics. Assume a continuous distribution of adsorbate energies ($\varepsilon$), corresponding to the density $\rho(\varepsilon)$. Note that $\rho(\varepsilon)$ is a sum of delta functions centered at given $\varepsilon$ multiplied by the ratio of the number of adsorption sites, normalized as

$$\int \rho(\varepsilon) d\varepsilon = 1. \tag{S9}$$

Since the Fermi-gas thermodynamic quantities are usually dependent on a chemical potential $\mu$, $\theta$ can be expressed for a monatomic adsorbed-phase Fermi gas as

$$\theta = \int \frac{1}{1+e^{(\varepsilon-\mu)/k_B T}} \rho(\varepsilon) d\varepsilon \tag{S10}$$

The vibrational and rotational contributions of the adsorbed phase ($E_{a,int}$), which may be applicable for diatomic molecules, are included in $\varepsilon$ as $\varepsilon = E_{ads} + E_{a,int}$, where $E_{a,int}$ is assumed as a constant. Since $E_{ads}$ becomes the quantity of interest, $\mu_a$ and $\theta$ can be expressed as

$$\mu_a = \mu - TS_{a,int} + E_{a,int} \tag{S11}$$

$$\theta = \int \frac{1}{1+e^{(E_{ads}-\mu)/k_B T}} \rho(E_{ads}) d(E_{ads}) \tag{S12}$$

Notably, $\mu$ only contains the configurational contribution, $S_{a,int}$ is the internal entropy (vibrational and rotational contributions), and $\rho(E_{ads})$ is the distribution of $E_{ads}$. For obtaining the desorption rate for a given $\rho(E_{ads})$ and $\theta$, Equation (S12) is solved numerically to acquire $\mu$. Then, by using Equations (S8), (S9), and (S11), the following expression is obtained

$$-\frac{d\theta}{dt} = \frac{sN_d p_0}{n_a \sqrt{2\pi m k_B T}} e^{N_d(\mu+E_{a,int})/k_B T} e^{-N_d S_{a,int}/k_B} e^{-\mu_{g0}/k_B T} \tag{S13}$$

for gases with tabulated values of $\mu_{g0}$. Equation (S13) is then integrated numerically to obtain $\theta(T)$, which can be used to obtain $\rho(E_{ads})$ from existing TPD spectra.[44]

### S7. Crystal orbital Hamilton population of isolated $O_2$

Crystal orbital Hamilton population (COHP) partitions the energy levels in the electronic structures into orbital-pair interactions, indicating bonding and antibonding contributions to the electronic structures. We calculated projected COHP (pCOHP) using LOBSTER.[47–50]

Figure S4 shows the spin-polarized projected density of states (PDOS) of O 2p and pCOHP of O-O bond in the isolated $O_2$, as well as the corresponding integrated pCOHP (IpCOHP). Notably, the pCOHP and IpCOHP shapes for the different functionals are almost the same for the isolated $O_2$, as well as the adsorptions on Pt(111) and *anatase*-$TiO_2$(101).

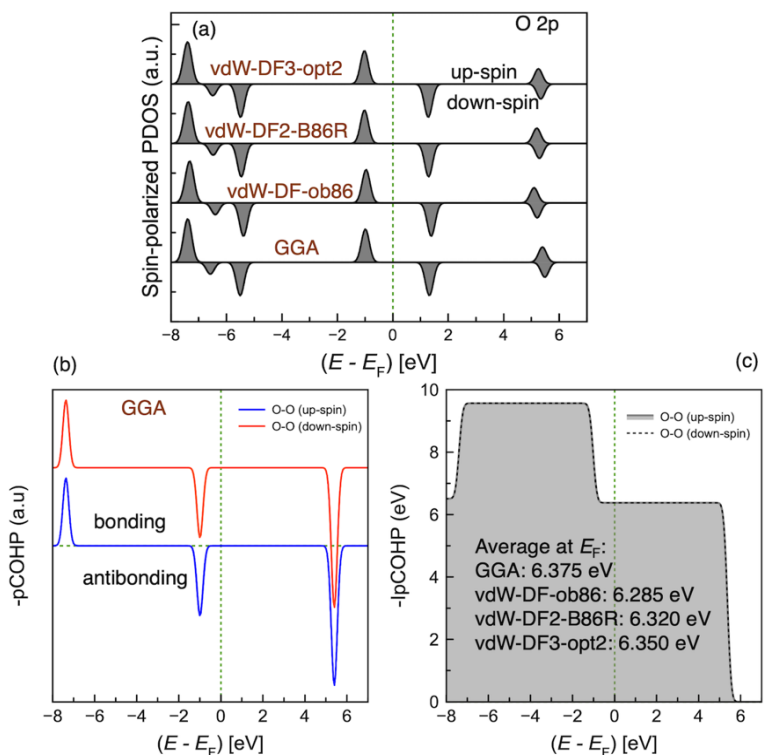

**Figure S4**. (a) Spin-polarized projected density of states (PDOS) of O 2p in the isolated $O_2$. (b) Projected crystal orbital overlap population (pCOHP) of O-O bond in the isolated $O_2$ for GGA as a sample. The occupied antibonding states indicate two unpaired spin-parallel electrons, accounting for $O_2$ paramagnetism. (c) Integrated projected crystal orbital overlap population (IpCOHP) of O-O bond in the isolated $O_2$. The average IpCOHP is also presented.

### S8. Crystal orbital Hamilton population for adsorptions of $O_2$ on Pt(111)

Figure S5 shows the single-spin total density of states (DOS) of Pt(111) as well as the projected DOS (PDOS) of Pt 5d states in the surface and bulk regions. Figure S5(b) shows the PDOS of $O_2$* 2p at the fcc site, as well as the PDOS of surface Pt atoms (Pt13, Pt15, and Pt16) of Pt(111) [see Figure S5(a)]. Figures S5(b) and S5(c) confirm that the metallic behavior of Pt(111) primarily stems from surface Pt 5d states at the Fermi level ($E_F$).

Figures S6(c) and S6(d) show pCOHP and IpCOHP, respectively, of bonds in $O_2$* and between $O_2$* and the surface Pt atoms. The numbering in Figure S6(a) is based on the *i*-th row in the input atomic position card. The metallic behavior is also confirmed in Figures S6(b) and S7(b). The contributions from O 2p states of $O_2$* and O* remain minimal and non-magnetized. Figure S6(d) indicates that the strong O-O bond within $O_2$* structure is maintained, as

S10

evidenced by a larger IpCOHP at $E_F$ for O17-O18 compared to Pt-O bonds (See Table S5, A). However, the value is less than that of the isolated $O_2$ depicted in Figure S4(c). As the average IpCOHP at $E_F$ for Pt-O bonds is not significantly influenced by the different functionals, the overestimated molecular $E_{ads}$ by vdW-DFs might not fully be correlated to the bonding characteristics.

Figure S7(b) shows the PDOS of O* 2p at the fcc site, as well as the PDOS of surface Pt atoms (Pt13, Pt15, and Pt16) of Pt(111), assigned in Figure S7(a). Figures S7(c) and S7(d) show pCOHP and IpCOHP of bonds between O* and the surface Pt atoms. The IpCOHP at $E_F$ of A, B, C, and D are presented in Table S5. The larger IpCOHP of Pt-O bond for vdW-DFs compared to GGA may partially correlate with the slight improvement in atomic $E_{ads}$. For both molecular and atomic adsorptions, the positive IpCOHPs confirm the bonding type characteristics.

Table S5. IpCOHP [in eV] at EF for $O_2$* and O* adsorptions on Pt(111)

| Functional | $O_2$* adsorption | | | | O* adsorption |
|---|---|---|---|---|---|
|  | A | B | C | D (average) | A |
| GGA | 5.271 | 1.426 | 0.909 | 1.082 | 1.412 |
| vdW-DF-ob86 | 5.096 | 1.437 | 0.936 | 1.103 | 1.413 |
| vdW-DF2-B86R | 5.090 | 1.441 | 0.940 | 1.107 | 1.413 |
| vdW-DF3-opt2 | 5.104 | 1.449 | 0.945 | 1.111 | 1.418 |

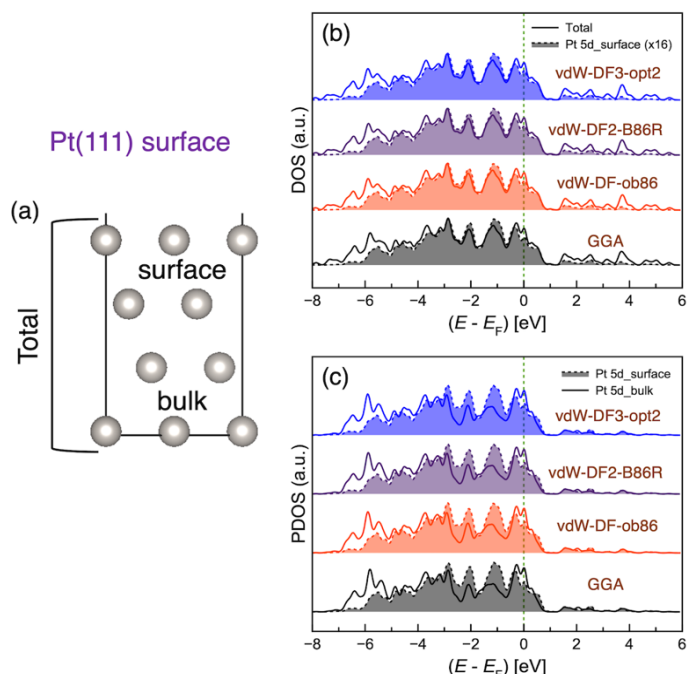

**Figure S5**. (a) Pt(111) structure showing surface and bulk regions with its (b) total DOS, as well as PDOS of surface Pt 5d states. (c) PDOS of bulk Pt 5d states is presented.

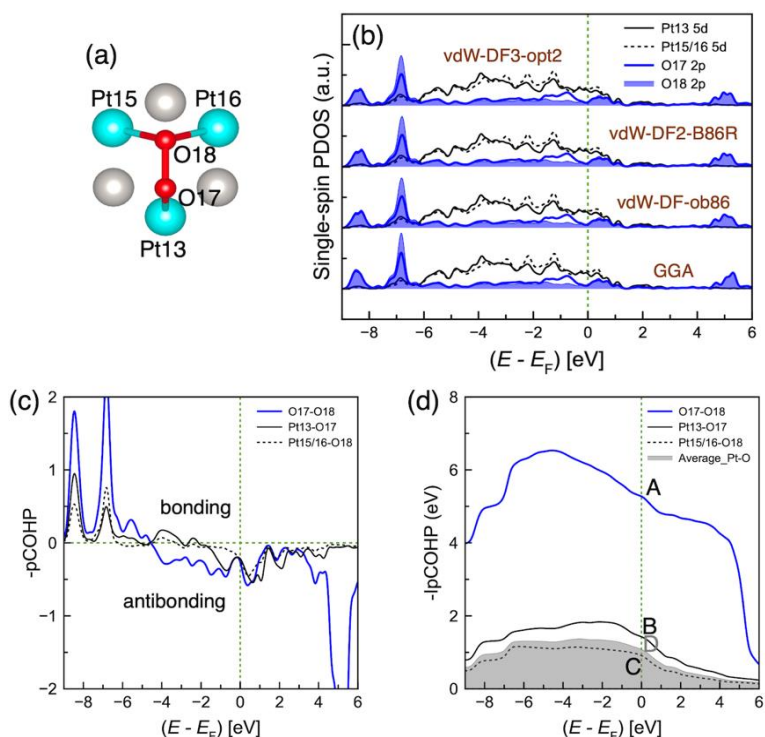

**Figure S6**. (a) Top view of surface Pt atoms of Pt(111) adsorbing $O_2^*$. (b) PDOS of surface Pt 5d and O 2p states of $O_2^*$ for different exchange-correlation functionals. (c) pCOHP and (d) IpCOHP) of bonds in $O_2^*$ and between $O_2^*$ and the surface Pt atoms.

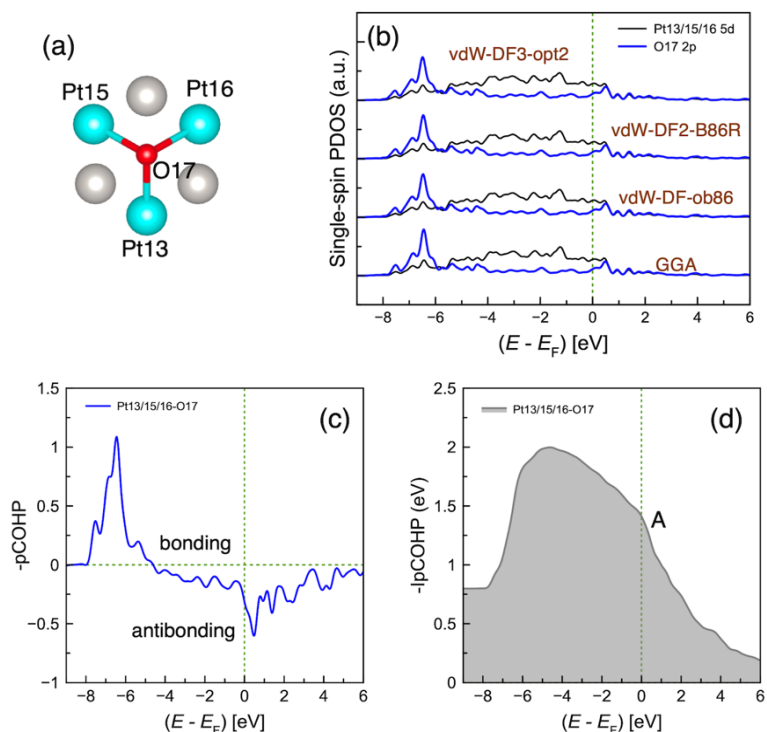

**Figure S7**. (a) Top view of surface Pt atoms of Pt(111) adsorbing $O^*$. (b) PDOS of the surface Pt 5d states and $O^*$ 2p states of $O^*$ for different exchange-correlation functionals. (c) pCOHP and (d) IpCOHP) of bonds between $O^*$ and the surface Pt atoms.

## S9. Crystal orbital Hamilton population for adsorptions of $O_2$ on *anatase*-$TiO_2$(101)

Figure S8 shows the total DOS of *anatase-TiO2*(101) for various $U$. Figure S9 shows PDOS of surface atoms, two Ti (Ti1, Ti2), and three O atoms (O1, O2, and O3). As both figures show similar DOS for different $U$, the following figures are presented for $U = 4$ eV. Figure S8 and S9 illustrate the minor effect of the Hubbard $U$ correction on the electronic structures, despite an apparent band gap (*c*) opening. Figure S10(b) shows spin-polarized PDOS of Ti31 3d and O 2p states in $O_2$* (O97 and O98), assigned in Figure S10(a). Figures S10(c) and S10(d) show pCOHP and IpCOHP, respectively, of bonds in $O_2$* and between $O_2$* and Ti31. The asymmetrical up-spin and down-spin PDOS and pCOHP in Figure S10(b) and S10(c), respectively, indicate that physisorption on *anatase*-$TiO_2$(101) maintains the spin-polarized adsorbed triplet state of $^3O_2$*. Figure S10(d) also indicates that the strong O-O bond within the $O_2$* structure is maintained, from the large down-spin IpCOHP at $E_F$ for O97-O98 [See Table S6, A(2)]. Conversely, the relatively low molecular $E_{ads}$ on *anatase*-$TiO_2$(101) can be explained by the low IpCOHP of Ti31–O98 bonds in Figure S10(d).

Figure S11(b) shows spin-polarized PDOS of Ti30 3d and O 2p states in surface O atom (O96) and O* (O97), assigned in Figure S11(a). Figures S11(c) and S11(d) show pCOHP and IpCOHP, respectively, of bonds between O* and the surface atoms. The IpCOHP at $E_F$ of A, B, C, and D are presented in Table S6. Despite the finite IpCOHP of the Ti30-O97 bond in Figure S11(d), the absence of stable atomic adsorption could be correlated with the inert-like behavior of *anatase*-$TiO_2$(101), which is similar to that of inert Au(111), which also avoids the atomic adsorption.[51] The bond-type adsorption on Pt(111) is distinguished from the other type. The molecular and atomic adsorptions on *anatase*-$TiO_2$(101) represent a mixed type with a slight predominance of the bonding type, as indicated by the positive IpCOHPs.

Table S6. IpCOHP [in eV] at $E_F$ for $O_2$* and O* adsorptions on *anatase*-$TiO_2$(101)

| Functional | $O_2$* adsorption | | O* adsorption | | |
|---|---|---|---|---|---|
| | A | B | A | B | C (average) |
| GGA | not calculated | | 0.677 | 1.833 | 1.255 |
| vdW-DF-ob86 (1) | -1.122 | 0.282 | 0.423 | 1.882 | 1.152 |
| (2) | 2.187 | 0.191 | | | |
| vdW-DF2-B86R (1) | -1.141 | 0.227 | 0.427 | 1.857 | 1.142 |
| (2) | 2.198 | 0.174 | | | |
| vdW-DF3-opt2 (1) | -1.118 | 0.272 | 0.381 | 1.836 | 1.109 |
| (2) | 2.276 | 0.212 | | | |

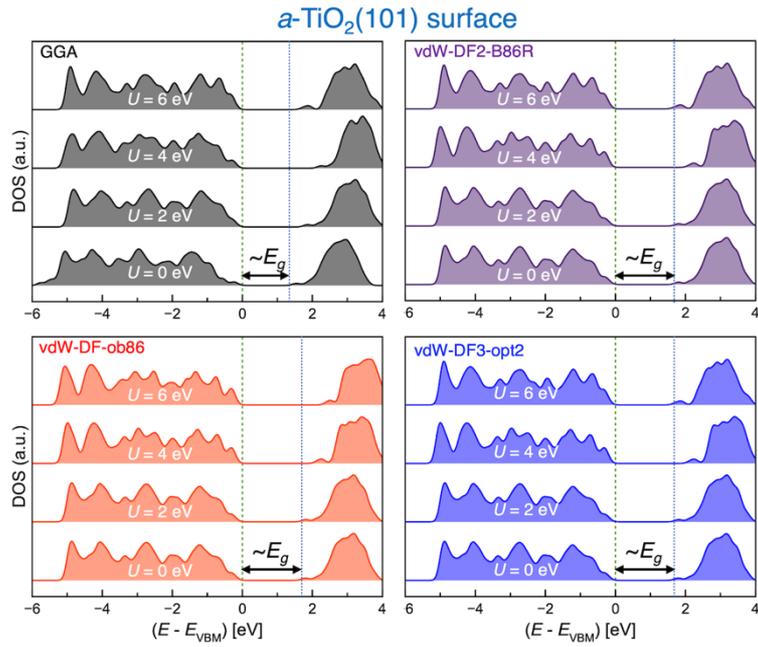

**Figure S8**. Total DOS of *anatase*-TiO$_2$(101) for different $U$.

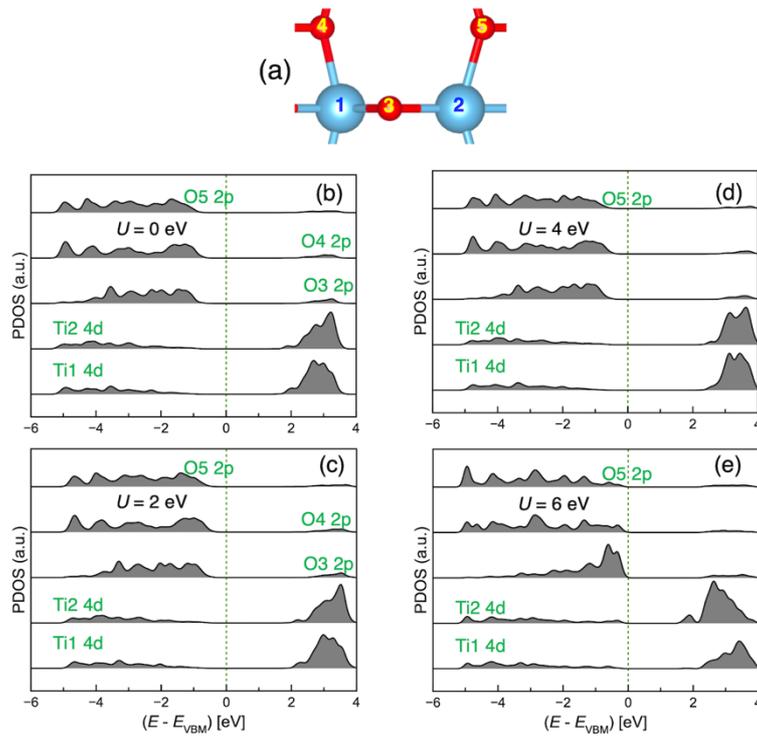

**Figure S9**. PDOS) of (a) surface atoms of *anatase*-TiO$_2$(101) for (b-e) different $U$ within GGA.

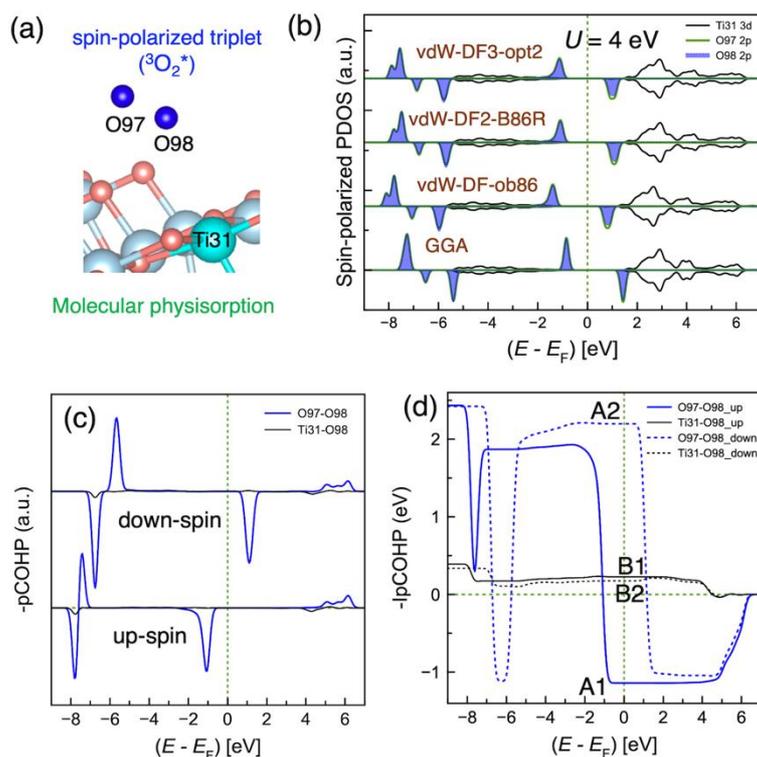

**Figure S10**. (a) Diagonal view of O$_2$* on *anatase*-TiO$_2$(101). (b) Spin-polarized PDOS of Ti31 3d and O 2p states of O$_2$* for $U = 4$ eV. (c) pCOHP and (d) IpCOHP) of bonds in O$_2$* and between O$_2$* and Ti 31.

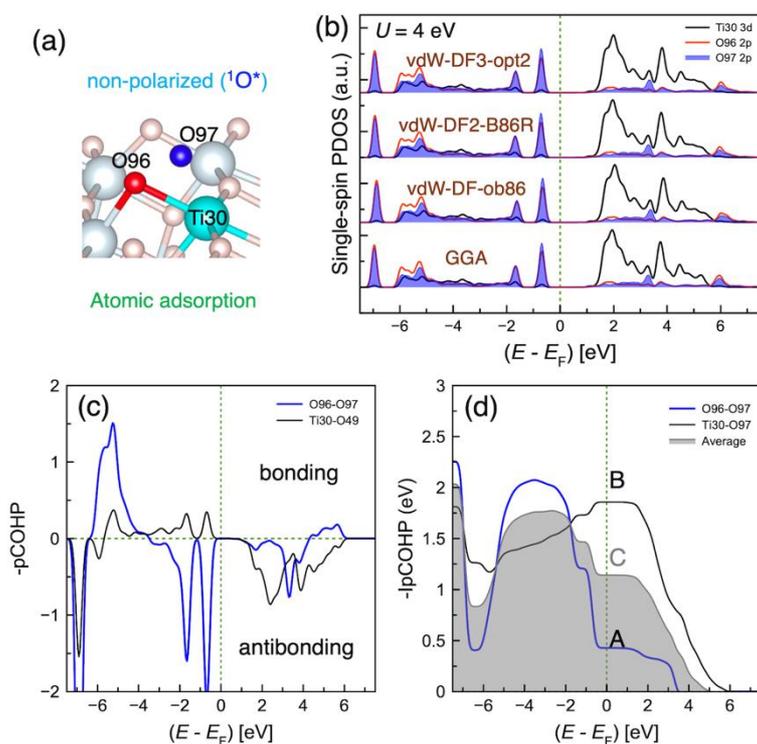

**Figure S11**. (a) Diagonal view of O* on *anatase*-TiO$_2$(101). (b) Spin-polarized PDOS of Ti30 3d, as well as O 2p states of O* and surface O atom (O96) for $U = 4$ eV. (c) pCOHP and (d) IpCOHP) of bonds in O$_2$* and between O$_2$* and Ti30.

**S10. Free energy calculation of oxygen reduction reaction on *anatase*-TiO$_2$(101).**

Each oxygen reduction reaction (ORR) elementary step in Eqs. (5)–(9) of the main text is correlated with a free energy change ($\Delta G$). $\Delta G$ contains contributions from changes in adsorption enthalpy ($\Delta H$) and entropic contribution ($T\Delta S$), expressed as

$$\Delta G = \Delta H - T\Delta S + \Delta G_\Phi + \Delta G_{pH} + \Delta G_{ef} + \Delta E_w \qquad (S14)$$

where $\Delta H$ is contributed by the energy change during the adsorption from DFT calculation ($\Delta E$) and zero-point energy ($\Delta E_{ZPE}$) related to vibrational frequency shift due to the adsorption, expressed as $\Delta H = \Delta E + \Delta E_{ZPE}$. Here, the $E_{ZPE}$ and $TS$ of the intermediates are adopted from the previous work.[27] $E_{ZPE}$ values of isolated molecules (H$_2$(*g*), O$_2$(*g*), and H$_2$O(*l*)) were calculated using phonon calculation. Furthermore, $TS$ terms of the isolated molecules were calculated by multiplying the standard entropy ($S°$) from NIST-JANAF database[25] by room temperature (298.15 K). $\Delta G_\Phi$ is the free energy change from the electrode potential ($\Phi$) as $\Delta G_\Phi = -e\Phi$. $\Delta G_{pH}$ and $\Delta G_{ef}$ are free energy changes related to the solution acidity (pH) and the electric double layer, respectively.[27] Notably, $\Delta G_{ef}$ can be included from the electric double layer since the surface charge densities at a fixed $\Phi$ are different for two surfaces.[52] However, we assume that this effect is negligible for the standard hydrogen electrode model. Moreover, because of the strong acidity limit (pH = 0), the effects of pH and the double electric layer can be ignored.[53] $\Delta E_w$ is the energy correction change from the solvation effect of environmental H$_2$O(*l*). The solvation correction energies for the intermediates are adopted from the previous work.[27] For obtaining the formation free energy of H$_2$O(*l*) at room temperature, which is $\Delta G_{H2O(l)} = -1.229$ eV ($-285.83$ kJ/mol),[25] $H$ of H$_2$(*g*) and O$_2$(*g*) are empirically corrected by $E_{OB}$ at 0 K and [$H$(298.15 K) – $H°$(0 K)], while $H$ of H$_2$O(*l*) is corrected by the empirical correction based on $\Delta_f H°$ at room temperature.[25]

From Eqs (5) to (9) and adopted from our previous work,[32] the 4e$^-$-transfer relative free energy of each reaction step, plotted in the free energy diagram at $\Phi = 1.229$ V, can be expressed as follows

(i)  $G(O_2^*) = [-4 \times \Delta G_{H2O(l)} + 4\Delta G_\Phi] + [E(O_2^*) - E(O_2) - E(*)]$
    $+ [E_{ZPE}(O_2^*) - E_{ZPE}(O_2)] + TS(O_2) + E_w(O_2^*),$ (S15)

(ii) $G(O_2H^*) = [-4 \times \Delta G_{H2O(l)} + 3\Delta G_\Phi] + [E(O_2H^*) - E(*) - E(½H_2) - E(O_2)]$
    $+ [E_{ZPE}(O_2H^*) - E_{ZPE}(½H_2) - E_{ZPE}(O_2)]$
    $- [-TS(½H_2) - TS(O_2)] + E_w(O_2H^*),$ (S16)

(iii) $G(O^*) = [-4 \times \Delta G_{H2O(l)} + 2\Delta G_\Phi] + [E(O^*) - E(*) + E(H_2O) - 2E(\frac{1}{2}H_2) - E(O_2)]$
$+ [E_{ZPE}(O^*) + E_{ZPE}(H_2O) - 2E_{ZPE}(\frac{1}{2}H_2) - E_{ZPE}(O_2)]$
$- [TS(H_2O) - 2TS(\frac{1}{2}H_2) - TS(O_2)] + E_w(O^*),$ (S17)

(iv) $G(OH^*) = [-4 \times \Delta G_{H2O(l)} + \Delta G_\Phi] + [E(OH^*) - E(*) + E(H_2O) - 3E(\frac{1}{2}H_2) - E(O_2)]$
$+ [E_{ZPE}(OH^*) + E_{ZPE}(H_2O) - 3E_{ZPE}(\frac{1}{2}H_2) - E_{ZPE}(O_2)]$
$- [TS(H_2O) - 3TS(\frac{1}{2}H_2) - TS(O_2)] + E_w(OH^*),$ (S18)

(v) $G(H_2O) = [-4 \times \Delta G_{H2O(l)}] + [2E(H_2O) - 4E(\frac{1}{2}H_2) - E(O_2)]$
$+ [2E_{ZPE}(H_2O) - 4E_{ZPE}(\frac{1}{2}H_2) - E_{ZPE}(O_2)]$
$- [2TS(H_2O) - 4TS(\frac{1}{2}H_2) - TS(O_2)].$ (S19).

Figures S12-S15 show the relaxed structures of intermediates alongside the corresponding $\Delta G$ at $\Phi = 0$ eV, calculated using vdW-DF3-opt2 + $U$. The similar structures are categorized with their similar $\Delta G$s. The numbers (1) to (12) assign the initial site number depicted in Figure S3.

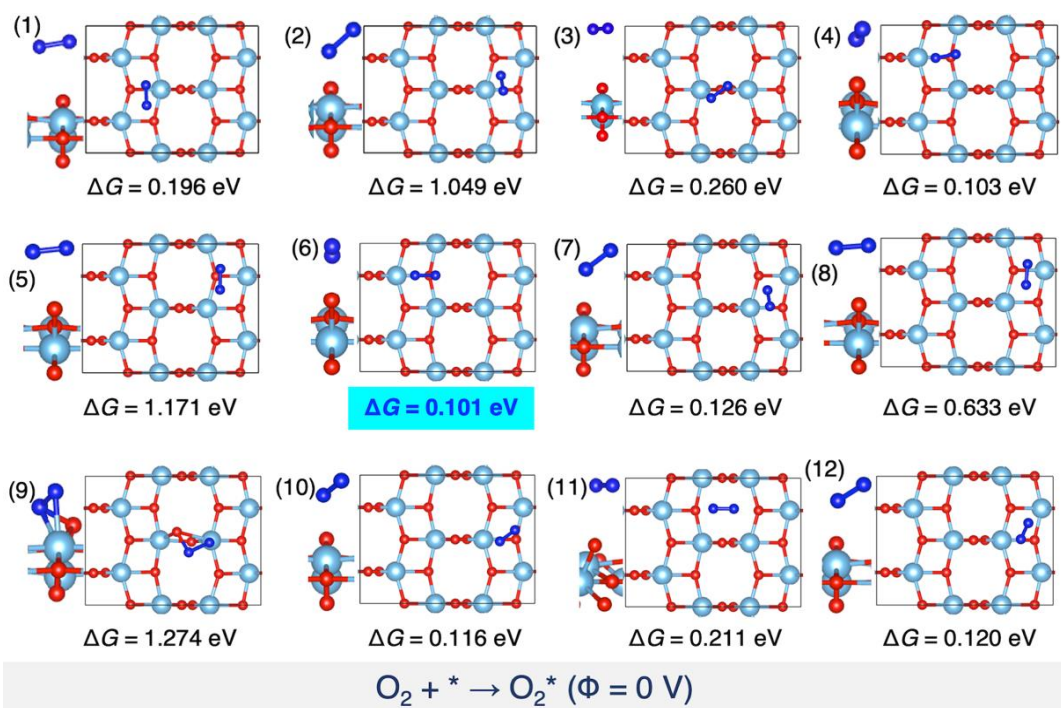

**Figure S12**. Relaxed structures of O$_2$* on *anatase*-TiO$_2$(101) with the corresponding $\Delta G$.

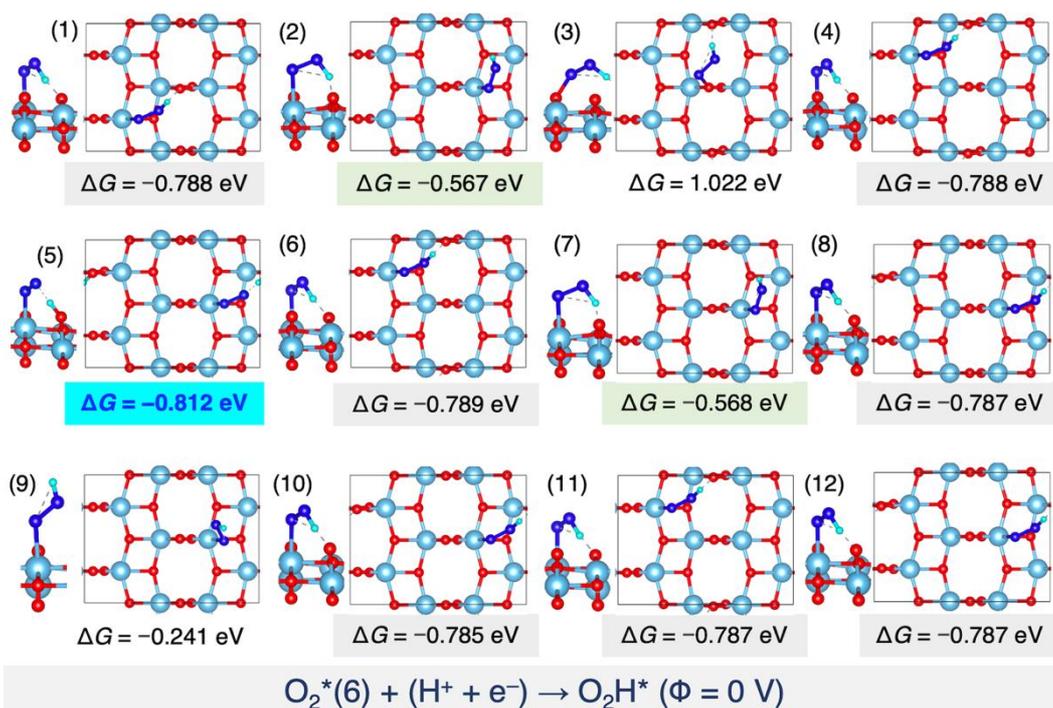

**Figure S13**. Relaxed structures of O$_2$H* on *anatase*-TiO$_2$(101) with the corresponding ΔG.

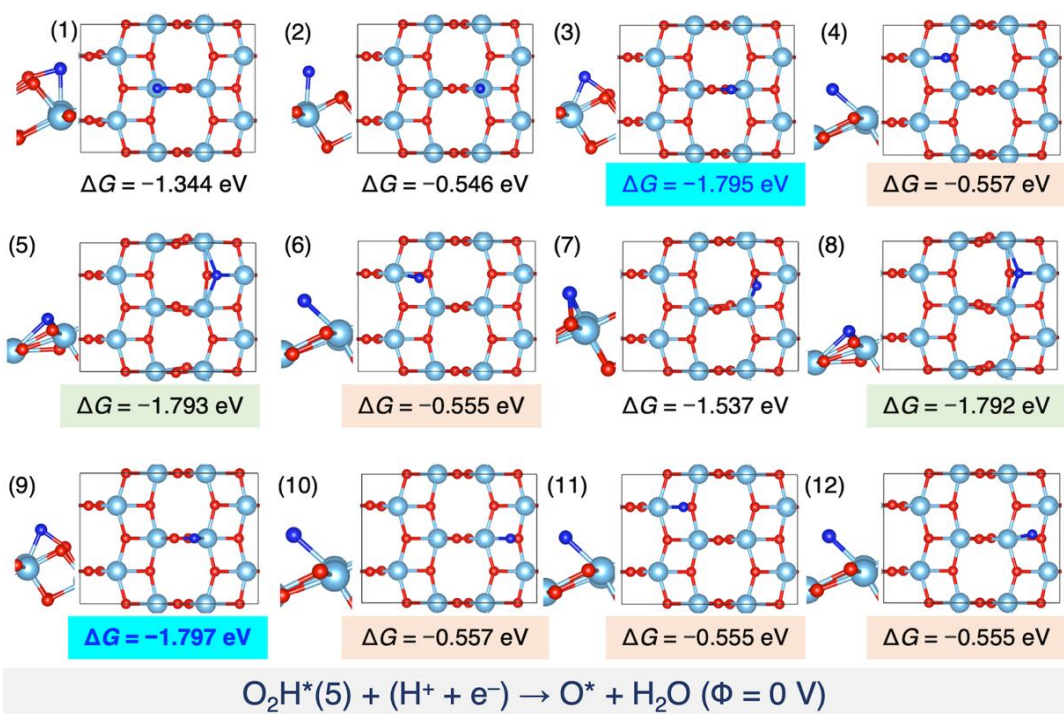

**Figure S14**. Relaxed structures of O* on *anatase*-TiO$_2$(101) with the corresponding ΔG.

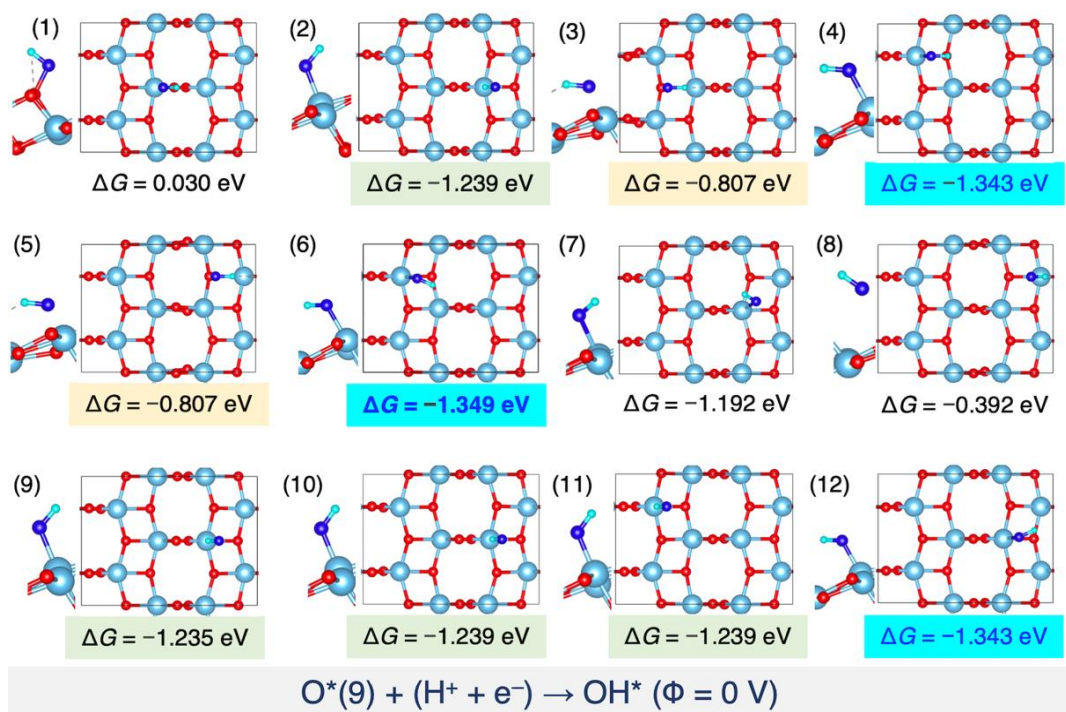

**Figure S15**. Relaxed structures of OH* on *anatase*-TiO$_2$(101) with the corresponding $\Delta G$.